\documentclass[aps,pre,nofootinbib,amsmath,amsfonts]{revtex4}
\usepackage{hyperref}
\usepackage{graphicx,xcolor}
\usepackage{longtable}

\newcommand{\eqr}[1]{Eq.~\eqref{#1}}

\newcommand{\secr}[1]{Sec.~[\ref{#1}]}
\newcommand{\ssecr}[1]{Subsec.~[\ref{#1}]}
\newcommand{\appr}[1]{Appendix~[\ref{#1}]}
\newcommand{\figr}[1]{Fig.~\ref{#1}}
\newcommand{\figsr}[2]{Figs.~\ref{#1}-\ref{#2}}

\newcommand{\velocity}{{\rm{v}}}
\newcommand{\vvelocity}{{\bf{v}}}
\newcommand{\vJ}{{\textbf{J}}}
\newcommand{\vR}{{\mathbf{r}}}
\newcommand{\vRt}{\vR,\,t}
\newcommand{\vg}{{\textbf{g}}}
\newcommand{\Kappa}{{\cal K}}
\newcommand{\nrange}{\overline{1,n}}
\newcommand{\nmorange}{\overline{1,n\!-\!1}}
\newcommand{\arglist}[2]{#1_{1}, \, \ldots, \, #1_{#2}}
\newcommand{\arglistr}[2]{#1_{1}(\vR), \, \ldots, \, #1_{#2}(\vR)}
\newcommand{\arglistrt}[2]{#1_{1}(\vRt), \, \ldots, \, #1_{#2}(\vRt)}

\newcommand{\alTcxi}{T, \, c, \, \xi}
\newcommand{\alTxi}{T, \, \xi}

\newcommand{\alnablarhoxi}{\rho,\,\xi^{m},\,\nabla{\rho},\,\nabla{\xi^{m}}}

\newcommand{\spd}{\!\cdot\!}
\newcommand{\PCDF}[2]{{\frac{\partial #1}{\partial x_{#2}}}} 
\newcommand{\SumVarea}{\vphantom{\sum\limits_1^1}}

\newcommand{\bigVarea}{\vphantom{\big(\big)}}

\newcommand{\vn}{{\mathbf{n}}}

\newcommand{\traceless}[1]{\mathring{#1}}

\newcommand{\tS}{{\,\mathrm{s}}}
\newcommand{\tV}{{\,\mathrm{v}}}
\newcommand{\tT}{{\,\mathrm{t}}}
\newcommand{\PPi}{\mathit\Pi}
\newcommand{\vpi}{\mathbf\pi}
\newcommand{\Tr}{\mathrm{Tr}}
\newcommand{\vdW}{{_{W}}}
\newcommand{\SRK}{{_{SRK}}}
\newcommand{\PR}{{_{PR}}}

\newcommand{\tsup}[1]{\textsuperscript{#1}}
\newcommand{\scaleprofile}{0.75}

\graphicspath{{figures/}}

\begin{document}

\title{The square gradient model in a two-phase mixture II. \\Non-equilibrium properties of a 2D-isotropic interface.}
\author{K.~S.~Glavatskiy}
\author{D.~Bedeaux}
\affiliation{Department of Chemistry, NO 7491, Norwegian University of Science and Technology, Trondheim, Norway}
\date\today

\begin{abstract}
In earlier work  \cite{bedeaux/vdW/I, bedeaux/vdW/II, bedeaux/vdW/III} a systematic extension of the van der Waals square gradient model to non-equilibrium one-component systems was given. In
this work the focus was on heat and mass transfer through the liquid-vapor interface as caused by a temperature difference or an over or under pressure. It was established that the surface as
described using Gibbs excess densities was in local equilibrium. Heat and mass transfer coefficients were evaluated. In our first paper \cite{glav/gradient/eq/I} we discussed the equilibrium
properties of a multi-component system following the same procedure. In particular, we derived an explicit expression for the pressure tensor and discussed the validity of the Gibbs relation in
the interfacial region. In this paper we will give an extension of this approach to multi-component non-equilibrium systems in the systematic context of non-equilibrium thermodynamics. The
two-dimensional isotropy of the interface is discussed. Furthermore we give numerically obtained profiles of the concentration, the mole fraction and the temperature, which illustrate the
solution for some special cases.
\end{abstract}

\maketitle


\numberwithin{equation}{section}

\section{Introduction.}

Because of the lack of local equilibrium the extension of non-equilibrium thermodynamics to a continuous description of an interface is not straightforward. In earlier work \cite{bedeaux/vdW/I,
bedeaux/vdW/II, bedeaux/vdW/III}, coauthored by one of us (DB), we were able to show that such an extension was possible for one-component fluids, with all the variables dependent on the normal
coordinate for a planar interface. Temperature gradients, pressure differences and the resulting heat flux and evaporation or condensation fluxes were determined through the interface. For
systems away from equilibrium, square gradient models have been used before, we refer to \cite{hohenberg/rmp, halperin/prb, penrose/pd} in this context. Very little work has been done on systems
with a varying temperature \cite{penrose/pd}, and on two-phase systems, however. The systematic treatment of heat and mass transport through the liquid-vapor interface, along the lines sketched
in \cite{bedeaux/vdW/I, bedeaux/vdW/II, bedeaux/vdW/III}, was to our knowledge, new. In the first paper \cite{glav/gradient/eq/I} we extended the analysis to equilibrium multi-component systems.
Explicit expression for all the thermodynamic quantities were given. In particular, we derived the expression for the pressure tensor and discussed the validity of the Gibbs relation in the
interfacial region. In this paper we will extend this approach to multi-component non-equilibrium mixtures in three-dimensions. Because of considering three-dimensional systems it is possible to
derive systematically all system properties, which were postulated in the one-dimensional description.

The traditional local equilibrium hypothesis implies that in non-equilibrium a small volume of the system at each moment of time can be considered to be in equilibrium. Thus, all thermodynamic
relations, valid for the whole system, remain valid for this small volume. The important assumption is that the state of each small volume is determined only by the properties of this volume,
and no other ones. To describe the interface in equilibrium, one needs to introduce a dependence on the density gradients however. Such a description is not local in the traditional sense: the
system behavior in a small volume depends on the properties of this volume and the properties of the nearest neighborhood. This implies that one cannot apply the usual local equilibrium
hypothesis to the interfacial region.

We will follow the traditional procedure used in non-equilibrium thermodynamics. In \secr{sec/Gradient} we review briefly the main results of the equilibrium square gradient model for the
interface. We extend all equilibrium results, derived in \cite{glav/gradient/eq/I} to non-equilibrium. Explicit expressions for the thermodynamic properties of the non-equilibrium mixture are
given. To be able to derive the entropy production one needs the Gibbs relation. In was shown in \cite{glav/gradient/eq/I} that within the interfacial region one can speak about two kinds of the
Gibbs relation, the ordinary Gibbs relation and the spatial Gibbs relation. We show in \secr{sec/Gradient} how these two static relations are combined to the non-equilibrium Gibbs relation. For
a one-component system the non-equilibrium Gibbs relation reduces to the one, given in \cite{bedeaux/vdW/I}. Within the framework of the one-fluid approach we discuss the expression for the
homogeneous Helmholtz energy $f_{0}$ which is needed to determine thermodynamic quantities in \secr{sec/Helmholtz/}. In \secr{sec/Hydrodynamics} we give the hydrodynamic equations. We use the
so-called one-fluid approach which has been shown \cite{henderson/2Fluid/II, brandani/2Fluid/III} to be appropriate for common mixtures. In \secr{sec/Curie2D} we discuss the consequences of the
special surface symmetry. After deriving the entropy production and using the Curie principle, we give the linear relations between the thermodynamic fluxes and forces in
\secr{sec/Hydrodynamics/PhenomenologicalEq}. It is found that for instance the resistances for transport through and into the interfacial region will in general contain square gradient
contributions. In \secr{sec/Profiles} we give for some examples numerically obtained concentrations, mole fraction and temperature profiles for a binary mixture with stationary mass and heat
transport through the interfacial region. Concluding remarks and a discussion are given in the last section.


%
\section{The square gradient model.}\label{sec/Gradient}
\subsection{Equilibrium surface.}\label{sec/Gradient/Equilibrium}

A mixture can be described by mass densities and temperature $T$ as independent variables. We will use $\rho(\vR)$ as the total mass density of the mixture, or $v^{m}(\vR) = 1/\rho(\vR)$ as the
mass specific volume, and $\{\arglistr{\xi^{m}}{n-1}\}$ as the mass fractions of components. Furthermore we will write $\xi^{m}$ instead of the set of arguments $\{\arglist{\xi^{m}}{n-1}\}$ and
$\nabla{\xi^{m}}$ instead of the set $\{\arglist{\nabla{\xi}^{m}}{n-1}\}$ to simplify the formulas.

Assuming that the specific Helmholtz energy can be written as
\begin{equation}\label{eq/Gradient/Eq/01}
f^{m}(\vR) = f^{m}_{0}(T,\,\rho,\,\xi^{m}) + \Kappa^{m}(\alnablarhoxi)
\end{equation}%
where
\begin{equation} \label{eq/Gradient/Eq/02}
\Kappa^{m}(\alnablarhoxi) \equiv  {\frac{1}{2}}{\frac{\kappa^{m}}{\rho(\vR)}}\,|\nabla{\rho(\vR)}|^{2} +
{\sum\limits_{i=1}^{n-1}{{\frac{\kappa_{i}^{m}}{\rho(\vR)}}\,\nabla{\rho(\vR)}\spd\nabla{\xi_{i}^{m}(\vR)}}} +
{\frac{1}{2}}{\sum\limits_{i,j=1}^{n-1}{{\frac{\kappa_{ij}^{m}}{\rho(\vR)}}\,\nabla{\xi_{i}^{m}(\vR)}\spd\nabla{\xi_{j}^{m}(\vR)}}} %
\end{equation}
and where all coefficients $\kappa$, $\kappa_{i}$ and $\kappa_{ij}$ are assumed to be independent of the temperature, one can derive the chemical potential $\mu_{n}$ of the
$n$\textsuperscript{th} component and the reduced chemical potential $\psi_{k} = \mu_{k} - \mu_{n}$, where $k=\nmorange$ (all integers from $1$ to $n-1$), which are constant through the surface
in equilibrium. We refer to \cite{glav/gradient/eq/I} for the expressions for the chemical potentials, $p(\vR)$ and other quantities. These quantities are related by the ordinary thermodynamic
relation
\begin{equation}\label{eq/Gradient/Eq/07}
f^{m}(\vR) = \mu_{n}^{m} - p\,(\vR)\,v^{m}(\vR) + {\sum\limits_{i=1}^{n-1}{\psi_{i}^{m}}}\xi_{i}^{m}(\vR)
\end{equation}%
Other thermodynamic potentials, like the specific internal energy
\begin{equation}\label{eq/Gradient/Eq/10}
u^{m}(\vR) = f^{m}(\vR) + s^{m}(\vR)\,T \\
\end{equation}
The entropy and the enthalpy are also given.

In the interfacial region, the pressure has a tensorial behavior:
\begin{equation}\label{eq/Gradient/Eq/08}
\sigma_{\alpha\beta}(\vR) = p\,(\vR)\,\delta_{\alpha\beta} + \gamma_{\alpha\beta}(\vR)
\end{equation}%
where the tension tensor is given by
\begin{equation}\label{eq/Gradient/Eq/09}
\gamma_{\alpha\beta}(\vR) = \kappa^{m}\,{\PCDF{\rho(\vR)}{\alpha}}{\PCDF{\rho(\vR)}{\beta}} +
{\sum\limits_{i=1}^{n-1}{\kappa_{i}^{m}\,\Big({\PCDF{\xi_{i}^{m}(\vR)}{\alpha}}{\PCDF{\rho(\vR)}{\beta}} + {\PCDF{\rho(\vR)}{\alpha}}{\PCDF{\xi_{i}^{m}(\vR)}{\beta}}\Big)}} +
{\sum\limits_{i,j=1}^{n-1}{\kappa_{ij}^{m}\,{\PCDF{\xi_{i}^{m}(\vR)}{\alpha}}{\PCDF{\xi_{j}^{m}(\vR)}{\beta}}}} %
\end{equation}%
For a flat surface $p(\vR)$ is the parallel pressure.

An important part of the equilibrium description is the relation between the rate of change of thermodynamic potentials and the independent thermodynamic variables, in other words the Gibbs
relation. In the interfacial region properties may change with the position, so we have to speak about two kinds of Gibbs relations: ordinary Gibbs relations and spatial Gibbs relations.
Ordinary Gibbs relations relate the changes of thermodynamic variables at the given point in space for different states. For the internal energy the ordinary Gibbs relation was found to be
\begin{equation}\label{eq/Gradient/Eq/11}
\delta u^{m}(s^{m}(\vR), v^{m}(\vR), \xi^{m}(\vR)) = T\,\delta s^{m}(\vR) + {\sum\limits_{i=1}^{n-1}{\psi^{m}_{i}\,\delta \xi^{m}_{i}(\vR)}} - p\,(\vR)\,\delta v^{m}(\vR)
\end{equation}%
Spatial Gibbs relations relate differences of thermodynamic variables for the given state at neighboring points in space. For the internal energy it was found to be
\begin{equation}\label{eq/Gradient/Eq/12}
\nabla u^{m}(\vR) = T\,\nabla s^{m}(\vR) + {\sum\limits_{i=1}^{n-1}{\psi_{i}^{m}\,\nabla \xi_{i}^{m}(\vR)}} - p\,(\vR)\,\nabla v^{m}(\vR) - v^{m}(\vR)\,{\PCDF{\gamma_{\alpha\beta}(\vR)}{\alpha}}
\end{equation}%
Notice in particular the last contribution on the right hand side.

\subsection{Non-equilibrium surface.}\label{sec/Gradient/NonEquilibrium}

In order to describe non-equilibrium processes in a multi-phase mixture using thermodynamics, one must assume, that all thermodynamic quantities are defined in each point in space and at all
times. This in particular is also in the interfacial region. In non-equilibrium the density $\rho(\vRt)$ and mass fractions $\xi^{m}(\vRt)$ depend on the time explicitly. We do not have the
restriction of a constant temperature and chemical potentials: $T(\vRt)$, $\mu_{n}^{m}(\vRt)$ and $\psi_{i}^{m}(\vRt)$ may therefore depend both on position and time. New variables which appear
in the non-equilibrium description are the velocities of each component $\{\arglistrt{\vvelocity}{n}\}$. We will however use the barycentric velocity of the whole mixture
\begin{equation}\label{eq/Gradient/NonEq/00}
\vvelocity(\vRt) = {\frac{1}{\rho(\vRt)}}{\sum\limits_{i=1}^{n}{\rho_{i}(\vRt)\vvelocity_{i}(\vRt)}} %
\end{equation}%
and the diffusion fluxes
\begin{equation}\label{eq/NonEquilibrium/DynamicLaws/03a}
{\vJ}_{k}^{m} \equiv \rho\,\xi_{k}^{m}\,(\vvelocity_{k}-\vvelocity), \qquad k=\nmorange
\end{equation}%
as independent variables.

To describe non-equilibrium inhomogeneous systems we shall assume that all the relations between thermodynamic variables valid in equilibrium, which were discussed in the previous section and in
more detail in \cite{glav/gradient/eq/I}, remain valid away from equilibrium. Away from the surface this is the usual assumption made in non-equilibrium thermodynamics. In the interfacial
regions it extends this assumption to places where the gradient contributions become important. As is said in \cite{deGrootMazur}, the validity of such a hypothesis can be verified only by
experiment.

An important part of the local equilibrium hypothesis is the relation between the rate of change of thermodynamic potentials and independent thermodynamic variables, in other words the Gibbs
relation. Similar to the description of a homogeneous fluid, we extend the equilibrium Gibbs relations in the simplest way. One needs to make one important observation before such an extension,
however: equilibrium equation \eqr{eq/Gradient/Eq/11} describe the change of local thermodynamic variables between two different states at a fixed point in space. These two states can be
separated in time. So we can say that this equation describes the change of local thermodynamic variables in time at a fixed point in space:
\begin{equation}\label{eq/Gradient/NonEq/08a}
T(\vRt)\,{\frac{\partial s^{m}(\vRt)}{\partial t}} = {\frac{\partial u^{m}(\vRt)}{\partial t}} - {\sum\limits_{i=1}^{n-1}{\psi _{i}^{m}(\vRt)\,{\frac{\partial\xi_{i}^{m}(\vRt)}{\partial t}}}} +
p(\vRt)\,{\frac{\partial v^{m}(\vRt)}{\partial t}}
\end{equation}%
We similarly use the equilibrium spatial Gibbs relation for the specific internal energy, \eqr{eq/Gradient/Eq/12}, for non-equilibrium case:
\begin{equation}\label{eq/Gradient/NonEq/08b}
T(\vRt)\,\nabla s^{m}(\vRt)  = \nabla u^{m}(\vRt) - {\sum\limits_{i=1}^{n-1}{\psi _{i}^{m}(\vRt)\,\nabla \xi_{i}^{m}(\vRt)}} + p(\vRt)\,\nabla v^{m}(\vRt) -
v^{m}(\vRt)\,{\PCDF{\gamma_{\alpha\beta}(\vRt)}{\alpha}}
\end{equation}%
See also relevant discussion in \appr{sec/Gibbs}. Contracting \eqr{eq/Gradient/NonEq/08b} with $\vvelocity(\vRt)$ and summing with \eqr{eq/Gradient/NonEq/08a} we find
\begin{equation}\label{eq/Gradient/NonEq/08}
T(\vRt)\,{\frac{ds^{m}(\vRt)}{dt}} = {\frac{du^{m}(\vRt)}{dt}} - {\sum\limits_{i=1}^{n-1}{\psi_{i}^{m}(\vRt)\,{\frac{d\xi_{i}^{m}(\vRt)}{dt}}}} + p(\vRt)\,{\frac{dv^{m}(\vRt)}{dt}} -
v^{m}(\vRt)\,\velocity_{\beta}(\vRt)\,{\PCDF{\gamma_{\alpha\beta}(\vRt)}{\alpha}}
\end{equation}%
where we used the substantial (barycentric) time derivative
\begin{equation}\label{eq/NonEquilibrium/DynamicLaws/02}
{\frac{d}{dt}} = {\frac{\partial }{\partial t}} + \vvelocity\spd\nabla %
\end{equation}%

\eqr{eq/Gradient/NonEq/08} is the Gibbs relation for the non-equilibrium two-phase mixture including the interface. One can show that it reduces to the Gibbs relation used by Bedeaux et al
\cite{bedeaux/vdW/I} for the case of a one-component fluid. The above analysis gives more insight in the origin of the contribution proportional to the divergence of the surface tension field.
This was not clarified in the analysis of Bedeaux et al. This defined all quantities and gave all relations needed in the non-equilibrium description using a generalization of the hypothesis of
local equilibrium. We will further omit the arguments $(\vRt)$ to simplify the notation.

\subsection{Mixing rules for the gradient coefficients.}\label{sec/Hydrodynamics/Coefficients}

Molar coefficients $\kappa^{m}$, $\kappa^{m}_{i}$ and $\kappa^{m}_{ij}$ are related to the volume coefficients $\kappa_{ij}^{v}$ as following (see \cite{glav/gradient/eq/I}):
\begin{equation}\label{}
\begin{array}{rl}
\SumVarea \kappa^{m} &= {\sum\limits_{i,j=1}^{n-1}{\xi_{i}^{m}\,\xi_{j}^{m}\,\mathrm{k}_{ij}^{v}}} + 2{\sum\limits_{i}^{n-1}{\xi_{i}^{m}\,\mathrm{k}_{i}^{v}}} + \mathrm{k}^{v}
\\
\SumVarea \kappa^{m}_{i} &= \rho\,{\sum\limits_{j=1}^{n-1}{\xi_{j}^{m}\,\mathrm{k}_{ij}^{v}}} + \rho\,\mathrm{k}_{i}^{v}
\\
\SumVarea \kappa^{m}_{ij} &= \rho^{2}\,\mathrm{k}_{ij}^{v}
\end{array}
\end{equation}
where
\begin{equation}\label{}
\begin{array}{rl}
\mathrm{k}_{ij}^{v} &\equiv \kappa_{ij}^{v}+\kappa_{nn}^{v}-\kappa_{in}^{v}-\kappa_{nj}^{v}
\\
\mathrm{k}_{i}^{v} &\equiv \kappa_{in}^{v}-\kappa_{nn}^{v}
\\
\mathrm{k}^{v} &\equiv \kappa_{nn}^{v}
\end{array}
\end{equation}
All these coefficients are in principle known functions of the densities. In practice they are not known for mixtures. Only the values for pure components are more or less known. Thus, it is
necessary to express the cross-coefficients in a form such that we can approximate them using pure-component values. From the equilibrium analysis for the specific quantities per unit of volume
\cite{glav/gradient/eq/I} one can see, that $\kappa_{ij}^{v}$ are simply related to the pure-component coefficients. We will assume then that, for instance, $\kappa_{ii}^{v}=\kappa_{i}^{v}$ is
the coefficient for the pure component $i$. Cross-coefficients can then be approximated by one of the so-called mixing rule for the gradient coefficients. We will assume the following mixing
rule
\begin{equation}\label{}
\kappa_{ij}^{v} = \sqrt{\kappa_{i}^{v}\kappa_{j}^{v}}
\end{equation}
This mixing rule for the volume coefficients is analogous to the mixing rule for $a_{ik}$ given in \secr{sec/Helmholtz/}.

\section{Homogeneous Helmholtz energy of one-fluid mixture.}\label{sec/Helmholtz/}

To obtain the homogeneous specific Helmholtz energy we use the most common one-fluid equations of state:

-- van der Waals equation of state:
\begin{equation}\label{eq/EOS/01}
p_{\vdW}(\alTcxi) = {\frac{RT}{v-B(\xi)}} - {\frac{A(\alTxi)}{v^2}} = {\frac{RTc}{1-B(\xi)\,c}} - A(\alTxi)\,c^{2}
\end{equation}

-- Soave-Redlich-Kwong equation of state:
\begin{equation}\label{eq/EOS/02}
p_{\SRK}(\alTcxi) = {\frac{RT}{v-B(\xi)}} - {\frac{A(\alTxi)}{v(v-B(\xi))}} = {\frac{RTc}{1-B(\xi)\,c}} - {\frac{A(\alTxi)}{1-B(\xi)\,c}}\,c^{2}
\end{equation}

-- Peng-Robinson equation of state:
\begin{equation}\label{eq/EOS/03}
p_{\PR}(\alTcxi) = {\frac{RT}{v-B(\xi)}} - {\frac{A(\alTxi)}{v^{2}+2B(\xi)\,v-B^{2}(\xi)}} = {\frac{RTc}{1-B(\xi)\,c}} - {\frac{A(\alTxi)}{1+2B(\xi)\,c - B^{2}(\xi)\,c^{2}}}\,c^{2}
\end{equation}
where $\xi$ is a short notation for the molar fractions $\{\arglist{\xi}{n}\}$.

In the one-fluid approach, constants $A(T)$ and $B$, depend on the fractions of the species due to the mixing rules:
\begin{equation}\label{eq/EOS/04}
\begin{array}{rl}
A(T,\, \arglist{\xi}{n}) &= {\sum\limits_{i,k = 1}^{n}{a_{ik}(T)\,\xi_{i}\,\xi_{k}}} \\
B(\arglist{\xi}{n}) &= {\sum\limits_{k = 1}^{n}{b_{k}\,\xi_{k}}}
\end{array}
\end{equation}
where usually $a_{ik}(T) = \sqrt{a_{i}(T)\,a_{k}(T)}$, and $a_{k}(T)$ and $b_{k}$ are the corresponding coefficients for pure substances. The matrix $a_{ik}(T)$ is symmetric in it's indexes.

To find the homogeneous specific Helmholtz energy $f_{0}(\alTcxi)$ we integrate the equations of state over the volume at constant temperature and molar fractions of the components. The
integration constant should be chosen such that the specific Helmholtz energy of the system with a small mixture's concentration ($c \rightarrow 0$) is equal to the specific Helmholtz energy for
a mixture of ideal gases. Integrating the equation of state and using \eqr{eq/IG/00} in \appr{sec/Helmholtz/Ideal}, we obtain following expression for the homogeneous specific Helmholtz energy:
\begin{equation}\label{eq/EOS/08}
f_{0}(\alTcxi) = -RT\,\ln\Big({\frac{e}{c\,N_{A}}}{\frac{\mathrm{w}(\alTxi)}{\Lambda^{3}(\alTxi)}}\,\big(1-B(\xi)\,c\big)\Big) - A(\alTxi)\,c\,\varphi\big(B(\xi)\,c\big)
\end{equation}
where $\varphi(\omega)$ has corresponding expression for each equation of state:
\begin{equation}\label{eq/EOS/09}
\begin{array}{rl}
\varphi_{\vdW}(\omega) &= 1 \\
\varphi_{\SRK}(\omega) &= {\frac{1}{\omega}}\ln(1 + \omega) \\
\varphi_{\PR}(\omega) &= {\frac{1}{2\sqrt{2}}}{\frac{1}{\omega}}\ln\big(1 + {\frac{2\sqrt{2}\,\omega}{1 + \omega(1-\sqrt{2})}}\big) \\
\end{array}
\end{equation}
\section{Hydrodynamics of one-fluid mixture.}\label{sec/Hydrodynamics}

We can now derive all hydrodynamic equations, using the conservation laws of matter, momentum and energy. The laws of conservation of mass can be written as
\begin{equation}\label{eq/NonEquilibrium/DynamicLaws/03}
\begin{array}{rl}
{\frac{d\rho }{dt}} &= -\rho\nabla\spd\vvelocity \\
\\
\rho {\frac{d\xi _{k}^{m}}{dt}} &= -\nabla\spd\vJ_{k}^{m}, \quad k=\nmorange
\end{array}
\end{equation}%

The momentum conservation law, or the equation of motion can be written as
\begin{equation}\label{eq/NonEquilibrium/DynamicLaws/06}
\rho \,{\frac{d\velocity_{\beta}}{dt}} = - {\PCDF{(\sigma_{\alpha\beta} + \pi_{\alpha\beta})}{\alpha}} + \rho g_{\beta}
\end{equation}%
where $\vg$ is the gravitational acceleration. $\sigma_{\alpha\beta}$ is the thermodynamic pressure tensor defined by \eqr{eq/Gradient/Eq/08} and $\pi_{\alpha\beta}(\vRt)$ is the viscous
pressure tensor, which still is to be determined. The viscous pressure tensor without subscripts will be indicated by $\Pi(\vRt)$. We assume, that this tensor is symmetric.

The law of energy conservation is (see \cite{deGrootMazur})
\begin{equation}\label{eq/NonEquilibrium/DynamicLaws/09}
\rho {\frac{de^{m}}{dt}} + \nabla\spd(\vJ_{e} - \rho \vvelocity e^{m})=0
\end{equation}%
where the total specific energy $e^{m}$ is given by
\begin{equation}\label{eq/Gradient/NonEq/04a}
e^{m}(\vRt) = u^{m}(\vRt) + \tau^{m}(\vRt) + \phi^{m}(\vR)
\end{equation}%
$\phi^{m}$ is the gravitational potential field, so that $\vg \equiv -\nabla\phi^{m}$. We will assume, that $\phi^{m}$ does not depend on the time.

We restrict ourself to systems, where the acceleration of the components relative to each other is small compared to the acceleration of the mixture's center of mass. This implies that the
kinetic energy of the components relative motion is small compared to the kinetic energy of the mixture's center of mass motion. This is true, when the relaxation time of the relative motion is
very small. For the common mixtures this is the case. Thus, the specific kinetic energy is
\begin{equation}\label{eq/Gradient/NonEq/04}
\tau^{m}(\vRt) = {\frac{1}{2}}\,\velocity^{2}(\vRt)
\end{equation}%
From momentum conservation we obtain:
\begin{equation}\label{eq/NonEquilibrium/DynamicLaws/07}
\rho \,{\frac{d\tau^{m}}{dt}} = -\velocity_{\beta}{\PCDF{(\sigma_{\alpha\beta} + \pi_{\alpha\beta})}{\alpha}} + \rho\vvelocity\spd\vg
\end{equation}%

For the internal energy we get
\begin{equation}\label{eq/NonEquilibrium/DynamicLaws/11}
\rho {\frac{du^{m}}{dt}} = - \nabla\spd\vJ_{q} - \pi_{\alpha\beta}\,\velocity_{\beta\alpha} - p\,\nabla\spd\vvelocity + \velocity_{\beta}\,{\PCDF{\gamma_{\alpha\beta}}{\alpha}}
\end{equation}%
where $\velocity_{\beta\alpha} \equiv {\partial\velocity_{\beta}/\partial x_{\alpha}}$ and where
\begin{equation}\label{eq/NonEquilibrium/DynamicLaws/13}
\vJ_{q} \equiv \vJ_{e} - \rho \,\vvelocity\,e^{m} - p\,\vvelocity - \Pi\spd\vvelocity %
\end{equation}%
is the total heat flux.

We write the Gibbs relation \eqr{eq/Gradient/NonEq/08} in the form
\begin{equation}\label{eq/NonEquilibrium/DynamicLaws/12}
T\,\rho \,{\frac{ds^{m}}{dt}} = \rho\,{\frac{du^{m}}{dt}} - {\sum\limits_{i=1}^{n-1}{\rho\,\psi_{i}^{m}\,{\frac{d\xi_{i}^{m}}{dt}}}} + p\rho \,\,{\frac{dv^{m}}{dt}} -
\velocity_{\beta}\,{\PCDF{\gamma_{\alpha\beta}}{\alpha}}
\end{equation}%
Using previous equations and performing algebraic transformations we obtain
\begin{equation}\label{eq/NonEquilibrium/DynamicLaws/14}
\rho \,{\frac{ds^{m}}{dt}} = -\nabla\spd{\frac{1}{T}}\,\Big( \vJ_{q} - {\sum\limits_{k=1}^{n-1}{\psi_{k}^{m}\,\vJ_{k}^{m}}} \Big) + \vJ_{q}\spd\nabla{\frac{1}{T}} -
{\sum\limits_{k=1}^{n-1}{\vJ_{k}^{m}\spd\nabla{\frac{\psi_{k}^{m}}{T}}}} - {\frac{1}{T}}\,\pi_{\alpha\beta}\,\velocity_{\beta\alpha}
\end{equation}
\section{2D isotropy of the surface.}\label{sec/Curie2D}

Even though the fluid does not have any preferred direction microscopically, we cannot say that it has a 3-dimensional isotropy everywhere, since there are the mesoscopic directions of the
density gradient. The two-phase equilibrium state is not 3-dimensionally isotropic.

A special care should be taken to determine the normal direction to the surface. With the help of the equilibrium analysis one can obtain the equilibrium densities distributions in the
interfacial region. It is possible therefore to determine the equidensity surfaces, i.e. mathematical surfaces, where either density is constant, and which are normal to the corresponding
density gradient. One may in principal use the gradients of either of the densities to define a direction normal to the surface. For the mixture we find it more convenient, however, to define a
normal using the tension field $\nabla_{\alpha}\gamma_{\alpha\beta}(\vR)$. We call the surfaces which are everywhere normal to this vector field the equitensional surfaces. The thickness of the
interfacial region will be assumed to be much smaller then the radii of curvature of these equitensional surfaces. Given this assumption the tension vector field in good approximation does not
change it's direction through the interface. Thus, it is possible to speak about the normal vector $\vn$ on the surface, which is parallel to the tension vector in this region.

This allows us to speak about the symmetry of the surface. If the surface curvature is the same in both directions, parallel to the surface, a small surrounding of the normal through the
interfacial region is invariant for any rotations around and reflections with respect to this normal. Thus we can say, that such a system has a local 2-dimensional isotropy. We shall refer to
such a property of the interfacial region as the 2-dimensional isotropy of the surface. If the two radii of curvature differ, surface is not 2-dimensionally isotropic any more. For a surface
which is thin compared to the radii of curvature one can, in a good approximation, consider it to be 2-dimensionally isotropic. We assume this to be the case for the systems we will consider.

If the system has 3-dimensional(3D) isotropy, then coupling occurs only between forces and fluxes of the same 3D tensorial character. For an interfacial region, which is 2-dimensionally(2D)
isotropic, coupling occurs only between forces and fluxes of the same 2D tensorial character. Thus, phenomenological coefficients must remain unchanged under rotations and reflections with
respect to the direction normal to the surface. Below we show how one can extract 2D-isotropic quantities from 3D scalars, vectors and tensors.

We shall use the special notation for the tensorial quantities of different order and different behavior in this section. Any tensorial quantity is denoted as $Q^{(d\,\mathrm{r})}$. Here $d$
indicates the dimensionality of the space, in which the quantity is being considered, and can be either 3 or 2 here. $\mathrm{r}$ indicates the rank of the tensorial quantity, and can be $\tS$
for scalar, $\tV$ for vectorial or $\tT$ for tensorial quantities. We refer to \appr{sec/2Din3D} for the details.

Consider the entropy production, which has a form
\begin{equation}\label{eq/NonEquilibrium/Curie/00}
\sigma_{s} = S^{(3\tS)}\,R^{(3\tS)} +  V^{(3\tV)} \spd W^{(3\tV)} +  T^{(3\tT)} : \PPi^{(3\tT)}
\end{equation}
To be able to use the 2D Curie principle one may proceed along the steps, explained in \cite{deGrootMazur}. To clarify this we shall write this expression as a combination of independent 2D
scalars, vectors and tensors. The details are given in \appr{sec/2Din3D}, here we will give the results.

One can split the vectorial and tensorial quantities into the normal and parallel components with respect to the normal vector $\vn$ on the surface. We use the subscripts $\perp$ and $\parallel$
for this quantities. Because of 2D-isotropy of the surface, these quantities reveal the scalar, vectorial or the tensorial behavior under rotations around and reflections with respect to this
normal in a 2D space (we refer to \appr{sec/2Din3D} for the details). This will be indicated by superscripts $2\,\mathrm{r}$ as explained above.

Any 3D scalar is also 2D scalar, since it remains invariant under any kind of coordinate transformations.
\begin{equation}\label{eq/NonEquilibrium/Curie/01}
S^{(3\tS)} = S^{(2\tS)} \equiv S^{(\tS)}
\end{equation}

Any 3D vector $V^(3\tV)$ can be written as (cf. the notation with \eqr{eq/NonEquilibrium/Curie/02}):
\begin{equation}\label{eq/NonEquilibrium/Curie/03a}
V^{(3\tV)} = \begin{pmatrix} V^{(\tS)}_{\perp},\,  V^{(2\tV)}_{\parallel} \end{pmatrix}
\end{equation}

Any 3D tensor $T^(3\tT)$ can be written as 
(cf. the notation with \eqr{eq/NonEquilibrium/Curie/07}):
\begin{equation}\label{eq/NonEquilibrium/Curie/08a}
\begin{array}{ll}
T^{(3\tT)} = \begin{pmatrix} T^{(\tS)}_{\perp\perp} & T^{(2\tV)}_{\perp\parallel} \\ T^{(2\tV)}_{\parallel\perp} & T^{(2\tT)}_{\parallel\parallel} \end{pmatrix} = %
\begin{pmatrix} T^{(\tS)}_{\perp\perp} & T^{(2\tV)}_{\perp\parallel} \\ T^{(2\tV)}_{\parallel\perp} & {\textstyle{1 \over 2}}\,(\Tr\,T^{(2\tT)}_{\parallel\parallel})\,U^{(2\tT)} + \traceless{T}^{(2\tT)}_{\parallel\parallel} \end{pmatrix} %
\end{array}
\end{equation}

Combining these components we obtain for the entropy production
\begin{equation}\label{eq/NonEquilibrium/Curie/09}
\sigma_{s} = \sigma_{s,scal} + \sigma_{s,vect} + \sigma _{s,tens}
\end{equation}
where
\begin{equation}\label{eq/NonEquilibrium/Curie/10}
\begin{array}{l}
\sigma_{s,\,scal} = S^{(\tS)}\,R^{(\tS)} + V^{(\tS)}_{\perp}\,W^{(\tS)}_{\perp} + T^{(\tS)}_{\perp\perp}\,\PPi^{(\tS)}_{\perp\perp} + {1 \over 2}(\Tr\,T^{(2\tT)}_{\parallel\parallel})\,(\Tr\,\PPi^{(2\tT)}_{\parallel\parallel}) \\
\\
\sigma_{s,\,vect} = V^{(2\tV)}_{\parallel} \spd W^{(2\tV)}_{\parallel} + 2\,T^{(2\tV)}_{\#} \spd \PPi^{(2\tV)}_{\#}\\
\\
\sigma _{s,\,tens} = \traceless{T}^{(2\tT)}_{\parallel\parallel} : \mathit{\traceless{\Pi}}^{(2\tT)}_{\parallel\parallel}%
\end{array}
\end{equation}
where $T^{(2\tT)}_{\#} \equiv {\textstyle{1 \over 2}}(T^{(2\tT)}_{\parallel\perp} + T^{(2\tT)}_{\perp\parallel})$ and we have used the symmetry of the tensor $T^{(3\tT)}$. The little circle
above a 2x2 tensor like in $\traceless{T}$ indicates the symmetric traceless part of this tensor. The 2D Curie principle tells us that coupling occur only between quantities of the same 2D
tensorial order.

\section{The phenomenological equations.}\label{sec/Hydrodynamics/PhenomenologicalEq}
\subsection{The force-flux relations.}\label{sec/Hydrodynamics/PhenomenologicalEq/ForceFlux}

Comparing \eqr{eq/NonEquilibrium/DynamicLaws/14} with the balance equation for the entropy
\begin{equation}\label{eq/NonEquilibrium/DynamicLaws/16}
\rho \,{\frac{ds^{m}}{dt}} =  -\nabla\spd\vJ_{s} + \sigma_{s}
\end{equation}%
we conclude, that the entropy flux and the rate of entropy production are given by
\begin{equation}\label{eq/NonEquilibrium/DynamicLaws/17}
\vJ_{s} = {\frac{1}{T}}\,\Big(\vJ_{q}-{\sum\limits_{k=1}^{n-1}{\psi _{k}^{m}\,\vJ_{k}^{m}}}\Big)
\end{equation}%
\begin{equation}\label{eq/NonEquilibrium/DynamicLaws/18}
\sigma_{s} = \vJ_{q}\spd\nabla{\frac{1}{T}} - {\sum\limits_{k=1}^{n-1}{{\vJ_{k}^{m}}\spd\nabla{\frac{\psi_{k}^{m}}{T}}}} - {\frac{1}{T}}\,\pi_{\alpha\beta}\,\velocity_{\beta\alpha}
\end{equation}
According to the second law $\sigma_{s}$ is non-negative. Comparing \eqr{eq/NonEquilibrium/DynamicLaws/18} with \eqr{eq/NonEquilibrium/Curie/09} and \eqr{eq/NonEquilibrium/Curie/10} we can write
the entropy production for a 2D-isotropic surface as the sum of 2-dimensional scalar, vectorial and tensorial contributions
\begin{equation}\label{eq/NonEquilibrium/PhenomenologicalEq/02}
\begin{array}{l}
\sigma_{s,\,scal} = J_{q,\,\perp}\,\nabla_{\perp}{\frac{1}{T}} - {\sum\limits_{k=1}^{n-1}{J_{i,\,\perp}^{\,m}\,\nabla_{\perp}{\frac{\psi_{k}^{m}}{T}}}} - (\nabla_{\perp}\velocity_{\perp})\,{1 \over T}\,\pi_{\perp\perp} - {1 \over 2}\,(\nabla_{\parallel}\spd\vvelocity_{\parallel})\,{1 \over T}\,(\Tr\,\pi_{\parallel\parallel}) \\
\\
\sigma_{s,\,vect} = \vJ_{q,\,\parallel}\spd\nabla_{\parallel}{\frac{1}{T}} - {\sum\limits_{k=1}^{n-1}{\vJ_{i,\,\parallel}^{m}\spd\nabla_{\parallel}{\frac{\psi_{k}^{m}}{T}}}} - 2\,\vvelocity_{\#}\spd{1 \over T}\,\vpi_{\#} \\
\\
\sigma _{s,\,tens} = (\traceless{\nabla_{\parallel}\,\vvelocity_{\parallel}}):{1 \over T}\,\traceless{\pi}_{\parallel\parallel}
\end{array}
\end{equation}
where $\vvelocity_{\#} \equiv {\textstyle{1 \over 2}}(\nabla_{\parallel}\velocity_{\perp} + \nabla_{\perp}\vvelocity_{\parallel})$.

The linear force-flux equations for the scalar force-flux pairs are
\begin{equation}\label{eq/NonEquilibrium/PhenomenologicalEq/05}
\begin{array}{rrrrrrrrr}
\nabla_{\perp}{\frac{1}{T}}                                 &=& R_{qq,\,\perp\perp}\,J_{q,\,\perp}          &-& {\sum\limits_{k=1}^{n-1}{R_{qk,\,\perp\perp}J_{k,\,\perp}^{\,m}}}          &-& R_{q\pi,\,\perp\perp}\,\pi_{\perp\perp}          &-& R_{q\pi,\,\perp\parallel}\,{1 \over 2}\,(\Tr\,\pi_{\parallel\parallel})  \\
\\
\nabla_{\perp}{\frac{\psi_{i}^{m}}{T}}                      &=& R_{iq,\,\perp\perp}\,J_{q,\,\perp}          &-& {\sum\limits_{k=1}^{n-1}{R_{ik,\,\perp\perp}J_{k,\,\perp}^{\,m}}}          &-& R_{i\pi,\,\perp\perp}\,\pi_{\perp\perp}          &-& R_{i\pi,\,\perp\parallel}\,{1 \over 2}\,(\Tr\,\pi_{\parallel\parallel})  \\
\\
{1 \over T}\,\nabla_{\perp}\velocity_{\perp}                &=& R_{\pi q,\,\perp\perp}\,J_{q,\,\perp}       &-& {\sum\limits_{k=1}^{n-1}{R_{\pi k,\,\perp\perp}J_{k,\,\perp}^{\,m}}}       &-& R_{\pi\pi,\,\perp\perp}\,\pi_{\perp\perp}        &-& R_{\pi\pi,\,\perp\parallel}\,{1 \over 2}\,(\Tr\,\pi_{\parallel\parallel})  \\
\\
{1 \over T}(\nabla_{\parallel}\spd\vvelocity_{\parallel})     &=& R_{\pi q,\,\parallel\perp}\,J_{q,\,\perp}   &-& {\sum\limits_{k=1}^{n-1}{R_{\pi k,\,\parallel\perp}J_{k,\,\perp}^{\,m}}}   &-& R_{\pi\pi,\,\parallel\perp}\,\pi_{\perp\perp}    &-& R_{\pi\pi,\,\parallel\parallel}\,{1 \over 2}\,(\Tr\,\pi_{\parallel\parallel}) %
\end{array}
\end{equation}
For the vectorial force-flux pairs they are
\begin{equation}\label{eq/NonEquilibrium/PhenomenologicalEq/06}
\begin{array}{rrrrrrr}
\vJ_{q,\,\parallel}     &=& L_{qq,\,\parallel\parallel}\,\nabla_{\parallel}{\frac{1}{T}}  &-& {\sum\limits_{k=1}^{n-1}{L_{qk,\,\parallel\parallel}\nabla_{\parallel}{\frac{\psi_{k}^{m}}{T}}}}   &-& L_{q\pi,\,\parallel\#}\,{1 \over T}\,\vvelocity_{\#}  \\
\\
\vJ_{i,\,\parallel}^{m} &=& L_{iq,\,\parallel\parallel}\,\nabla_{\parallel}{\frac{1}{T}}  &-& {\sum\limits_{k=1}^{n-1}{L_{ik,\,\parallel\parallel}\nabla_{\parallel}{\frac{\psi_{k}^{m}}{T}}}}   &-& L_{i\pi,\,\parallel\#}\,{1 \over T}\,\vvelocity_{\#}  \\
\\
2\vpi_{\#}         &=& L_{\pi q,\,\#\parallel}\,\nabla_{\parallel}{\frac{1}{T}}      &-& {\sum\limits_{k=1}^{n-1}{L_{\pi i,\,\#\parallel}\nabla_{\parallel}{\frac{\psi_{k}^{m}}{T}}}}       &-& L_{\pi\pi,\,\#\#}\,{1 \over T}\,\vvelocity_{\#}  \\
\end{array}
\end{equation}
and for the tensorial force-flux pairs they are
\begin{equation}\label{eq/NonEquilibrium/PhenomenologicalEq/08}
\begin{array}{rrr}
\traceless{\pi}_{\parallel\parallel} &=& L_{\pi}\,{1 \over T}\,(\traceless{\nabla_{\parallel}\,\vvelocity_{\parallel}})
\end{array}
\end{equation}
All the resistivities $R$ and conductivities $L$ are scalars. One can easily invert the resistivity matrix $R$ and write the corresponding relations for the fluxes using the conductivities $L$,
and vice versa.

For flat surfaces it was found \cite{BedeauxOptical} that the resistivities are additive in the normal direction to the surface while the conductivities are additive in the parallel direction.
We therefore consider it convenient to write the force-flux relations in the given form.

%
\subsection{The phenomenological coefficients.}\label{sec/Hydrodynamics/PhenomenologicalEq/Coefficients}

The Onsager relations for the phenomenological coefficients are the following.
\begin{equation}\label{eq/NonEquilibrium/PhenomenologicalEq/10}
\begin{array}{lllllll}
\bigVarea R_{qk,\,\perp\perp}           &=  &R_{kq,\,\perp\perp}            & \qquad\qquad &    \bigVarea L_{qk,\,\parallel\parallel}   &=  &L_{kq,\,\parallel\parallel} \\
\bigVarea R_{q\pi,\,\perp\perp}         &=  &R_{\pi q,\,\perp\perp}         & \qquad\qquad &    \bigVarea L_{q\pi,\,\parallel\#}        &=  &L_{\pi q,\,\#\parallel} \\
\bigVarea R_{q\pi,\,\perp\parallel}     &=  &R_{\pi q,\,\parallel\perp}     & \qquad\qquad &    \bigVarea L_{ik,\,\parallel\parallel}   &=  &L_{ki,\,\parallel\parallel} \\
\bigVarea R_{ik,\,\perp\perp}           &=  &R_{ki,\,\perp\perp}            & \qquad\qquad &    \bigVarea L_{i\pi,\,\parallel\#}        &=  &L_{\pi i,\,\#\parallel} \\
\bigVarea R_{i\pi,\,\perp\perp}         &=  &R_{\pi i,\,\perp\perp}         & \qquad\qquad &&&  \\
\bigVarea R_{i\pi,\,\perp\parallel}     &=  &R_{\pi i,\,\parallel\perp}     & \qquad\qquad &&&  \\
\bigVarea R_{\pi\pi,\,\perp\parallel}   &=  &R_{\pi\pi,\,\parallel\perp}    & \qquad\qquad &&&  \\
\end{array}
\end{equation}
As the ordinary Onsager relations, these are the consequence of the microscopic time reversal invariance.

As usual the values of the phenomenological coefficients locally will depend on the local thermodynamic variables. These are the local concentrations, $\rho$, $\xi^{m}$, and the temperature,
$T$. In the gradient theory the density gradients are also considered as local thermodynamic variables. In view of this the phenomenological coefficients may also depend on the gradients of the
densities. The values of the phenomenological coefficients  and their functional dependence on the thermodynamic variables are not given by the mesoscopic theory. They should be either
calculated from statistical mechanical considerations, or from experiments, either real or computer. While they are well investigated for homogeneous fluids and fluid mixtures, such data are not
available for the surface coefficients of fluid mixtures.

We shall use the following expression for each of the resistivity coefficients
\begin{equation}\label{eq/NonEquilibrium/PhenomenologicalEq/11}
R = R^{I} + (R^{II}-R^{I}){\frac{\varphi-\varphi^{I}}{\varphi^{II}-\varphi^{I}}} + \alpha\,(R^{II}+R^{I}){\frac{|\nabla{\varphi}|^{2}}{|\nabla{\varphi_{eq}}|_{\max}^{2}}}
\end{equation}
where $R^{I}$ and $R^{II}$ are the resistivities for the coexisting homogeneous phases in the equilibrium state. Here $\varphi$ is the order parameter, typically this is just the density $\rho$
or the molar concentration $c$. $\varphi_{eq}$ is the equilibrium profile and $|\nabla\varphi_{eq}|_{\max}$ is the maximum value of the gradient of this profile. The first two terms are just a
smooth transition of the resistivity from the value in the one phase to the value in the other phase. This is the first natural assumption for the resistivity profile. The origin of the third
term comes from the assumption of an excess resistivity in the interfacial region. Particularly one can observe this fact in the molecular dynamic simulations \cite{surfres}. The exact form of
this term may be debated. It was chosen to model a rise of the resistivity in the interfacial region. The $|\nabla{\varphi}|^{2}$ factor makes this term significant only in the interfacial
region. It is scaled with $|\nabla{\varphi_{eq}}|_{\max}^{2}$ in order to make this factor dimensionless and not far from unity.  The $(R^{II}+R^{I})$ factor gives the average value of the
resistivity of both phases. The dimensionless factor $\alpha$ determines the magnitude of this effect. The homogeneous resistivities $R^{I}$ and $R^{II}$ are the known functions of the mass
fraction and the temperature along the plane of coexistence.

For the conductivities used in \eqr{eq/NonEquilibrium/PhenomenologicalEq/06} and \eqr{eq/NonEquilibrium/PhenomenologicalEq/08} one may use expressions analogous to
\eqr{eq/NonEquilibrium/PhenomenologicalEq/11}. The conductivities along the surface are expected to be additive \cite{BedeauxOptical}. Thus it is important to use this equation for the
conductivities and not for the resistivities along the surface. In this respect it is important to note that $\alpha$ may in principle be negative as long as the corresponding $R$ and $L$ remain
everywhere positive. For the resistivity this would describe an interfacial region with a lower resistivity and for the conductivities it would describe an interfacial region with a lower
conductivity. Below we will only consider positive $\alpha$'s.

\section{Typical profiles for the binary mixture.}\label{sec/Profiles}

In order to illustrate the results, which one can obtain using the above procedure, we have applied it to a special case. This requires a number of approximations, connected with the specific
mixture and the geometry. We consider a flat liquid-vapor interface of the binary mixture of cyclohexane (1\tsup{st} component) and $n$-hexane (2\tsup{nd} component) in non-equilibrium
stationary conditions. We only consider fluxes and gradients in the direction normal to the surface. Furthermore we neglect viscous contributions. The force-flux relations then reduce to
\begin{equation}\label{eq/NonEquilibrium/Profiles/01}
\begin{array}{rrrrr}
\SumVarea {\frac{d}{dx}}{\frac{1}{T}}           &=& R_{qq}\,J_{q} &-& R_{q1}J_{1}^{\,m}\\
\SumVarea {\frac{d}{dx}}{\frac{\psi^{m}}{T}}    &=& R_{1q}\,J_{q} &-& R_{11}J_{1}^{\,m}\\
\end{array}
\end{equation}
The differential equations for the temperature, density and fraction profiles were solved using a numerical method for a two point boundary value problem. This was done using a collocation
method implemented in the Matlab function bvp4c \cite{bvp4c}. In this procedure we used the equilibrium profiles, found in \cite{glav/gradient/eq/I} as an initial state. Further details of the
solution procedure will be given in the following paper. The numerical values of the homogeneous resistivities were taken from \cite{Data/Transport/Dong}. In all cases the integrated molar
content for both components was kept equal to the equilibrium value.

The first aspect we will try to clarify is the influence of the additional resistivity to transport. We consider in particular two cases. In the first only $\alpha_{qq}$, the $\alpha$-factor for
the heat resistivity coefficient $R_{qq}$, is unequal to zero and in the second only $\alpha_{ii}$, the $\alpha$-factor for the diffusion resistivity coefficient $R_{ii}$, is unequal to zero. In
\figsr{Fig_c-Aqq-p=0,95pe_Tg=Tl=Te}{Fig_T-Aqq-p=0,95pe_Tg=Tl=Te} we plot the total molar concentration, the mole fraction and the temperature for the case that only $\alpha_{qq} \neq 0$. The
system is brought out of equilibrium by reducing the pressure on the vapor side to $0.95\,p_{eq}$, where $p_{eq}$ is the equilibrium pressure. The temperatures on both ends of the box are kept
equal to equilibrium temperature $T_{eq}$. In \figr{Fig_c-Aqq-p=0,95pe_Tg=Tl=Te} we see, that the total molar concentration does not depend on the value of $\alpha_{qq}$ very much. This is
different for the mole fraction which increases about 2 $\%$ on the vapor side when $\alpha_{qq}$ increases from 0 to 10. The temperature decreases due to the evaporation. In all cases the
extrapolated temperature in the liquid is higher then the value in the vapor, where we extrapolate to the inflection point of the total molar concentration. For $\alpha_{qq}=10$ the minimum of
the temperature is below both extrapolations. In that case the temperature "jump" in the the extrapolated profiles has increased to about 3 $^{\circ}$C.
%
%
\begin{figure}[hbt!]
\centering
\includegraphics[scale=\scaleprofile]{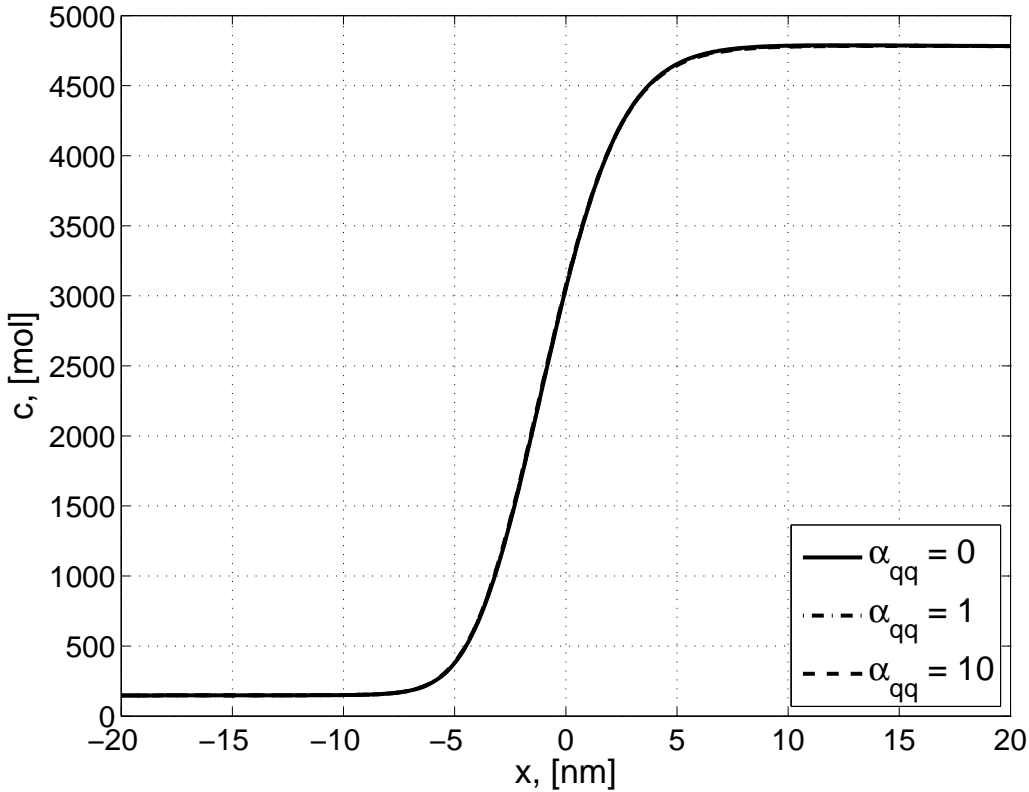}
\caption{Molar concentration profile for different $\alpha_{qq}$ at $p=0.95 p_{eq}$ and $T_{g}=T_{\ell}=T_{eq}$.}\label{Fig_c-Aqq-p=0,95pe_Tg=Tl=Te}
\end{figure}
\begin{figure}[hbt!]
\centering
\includegraphics[scale=\scaleprofile]{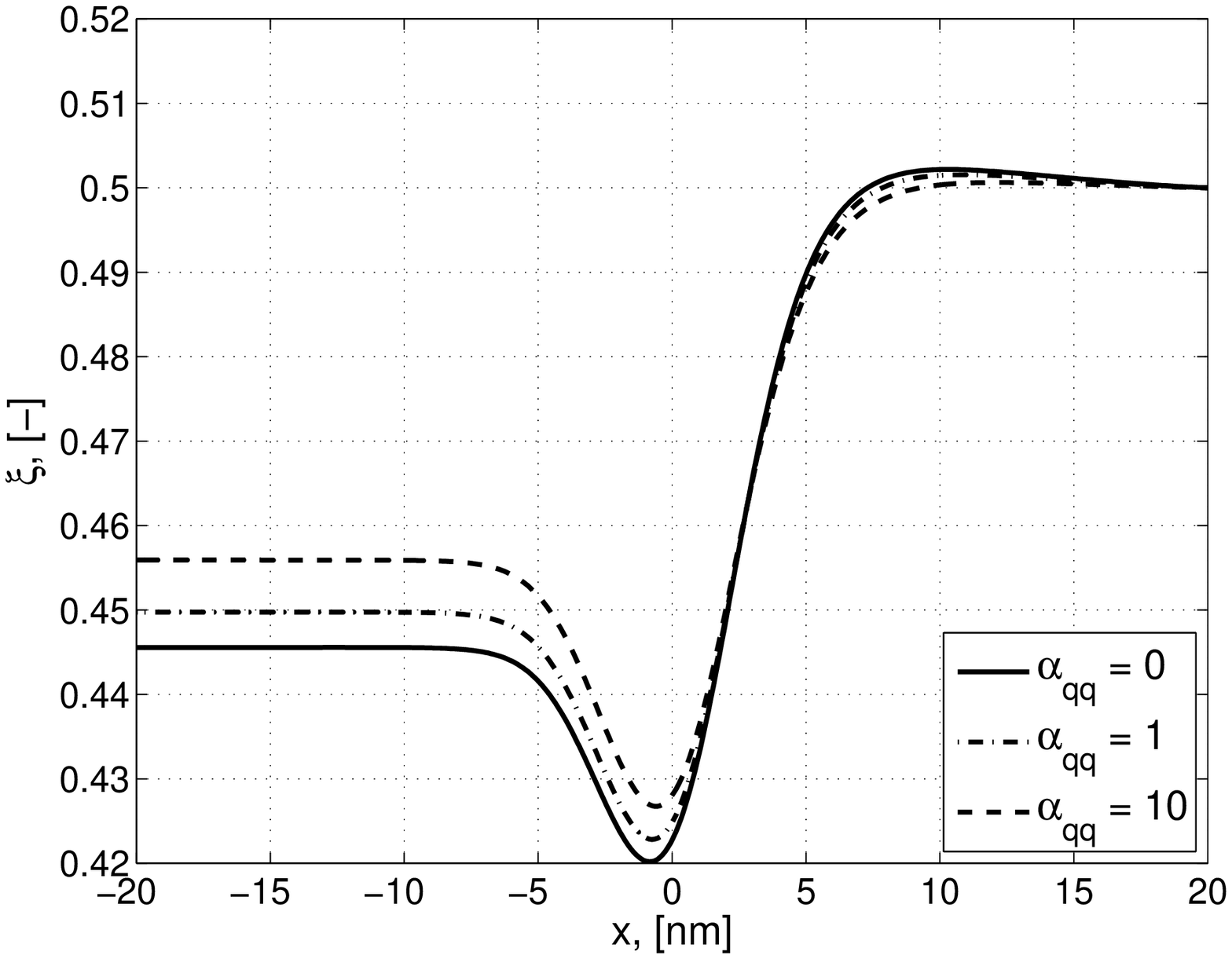}
\caption{Molar fraction profile for different $\alpha_{qq}$ at $p=0.95 p_{eq}$ and $T_{g}=T_{\ell}=T_{eq}$.}\label{Fig_xi-Aqq-p=0,95pe_Tg=Tl=Te}
\includegraphics[scale=\scaleprofile]{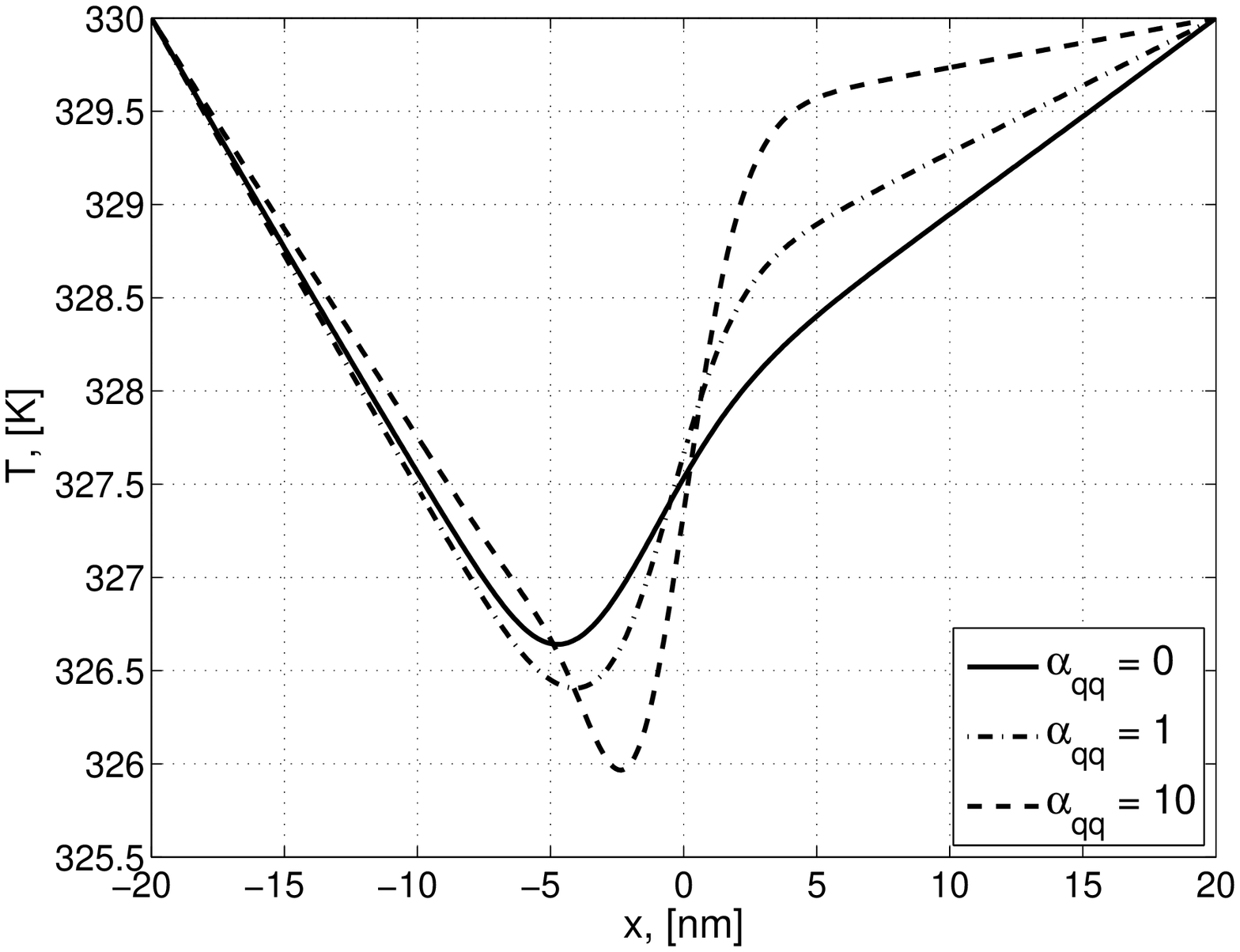}
\caption{Temperature profile for different $\alpha_{qq}$ at $p=0.95 p_{eq}$ and $T_{g}=T_{\ell}=T_{eq}$.}\label{Fig_T-Aqq-p=0,95pe_Tg=Tl=Te}
\end{figure}

In \figsr{Fig_xi-Aii-p=0,95pe_Tg=Tl=Te}{Fig_T-Aii-p=0,95pe_Tg=Tl=Te} we plot the mole fraction and the temperature for the case that only $\alpha_{ii} \neq 0$. Further conditions are the same as
in the previous example. We did not plot the total molar concentration which is not very different from the one given in \figr{Fig_c-Aqq-p=0,95pe_Tg=Tl=Te}. The modification of the mole fraction
is now more dramatic. In the vapor it decreases up to 27 $\%$. The temperature increases for larger values of $\alpha_{ii}$. This is related to a decrease of the evaporation. The temperature
jump in the extrapolated profiles is in all cases not more than 0.5 $^{\circ}$C.
%
%
\begin{figure}[hbt!]
\centering
\includegraphics[scale=\scaleprofile]{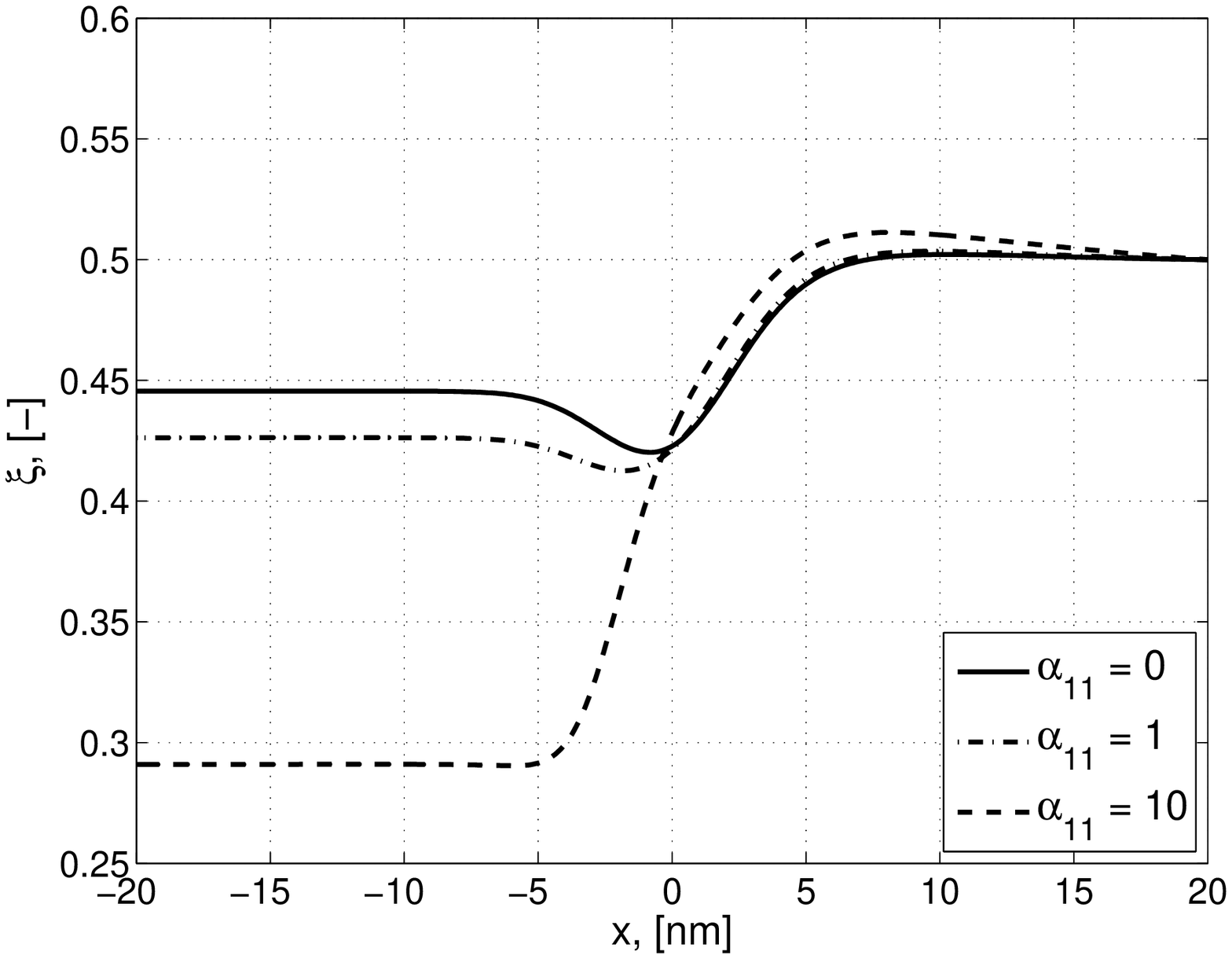}
\caption{Molar fraction profile for different $\alpha_{11}$ at $p=0.95 p_{eq}$}\label{Fig_xi-Aii-p=0,95pe_Tg=Tl=Te}
\includegraphics[scale=\scaleprofile]{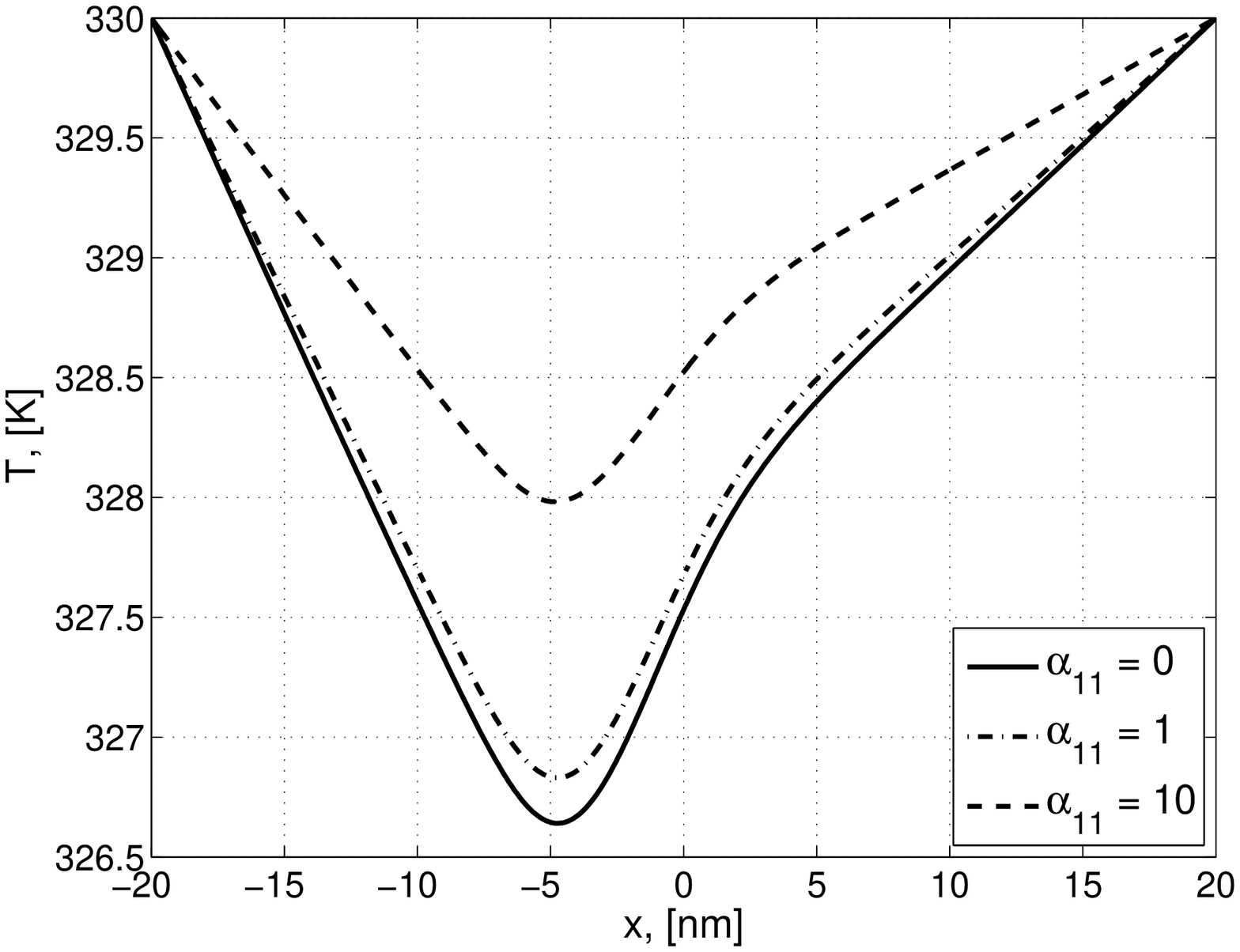}
\caption{Temperature profile for different $\alpha_{11}$ at $p=0.95 p_{eq}$}\label{Fig_T-Aii-p=0,95pe_Tg=Tl=Te}
\end{figure}

The third example considers the $\alpha_{qq} \neq 0$ case when the vapor pressure and temperature are kept equal to the equilibrium values and the liquid temperature is 5 $\%$ higher then the
equilibrium value. The total concentration profile is again similar to \figr{Fig_c-Aqq-p=0,95pe_Tg=Tl=Te} and not given. \figsr{Fig_xi-Aqq-p=pe_Tg=Te_Tl=1,05Te}{Fig_T-Aqq-p=pe_Tg=Te_Tl=1,05Te}
give the mole fraction and temperature profiles. The change in the mole fraction went up to 12 $\%$. The temperature jump goes up to about 20 $^{\circ}$C.
%
%
\begin{figure}[hbt!]
\centering
\includegraphics[scale=\scaleprofile]{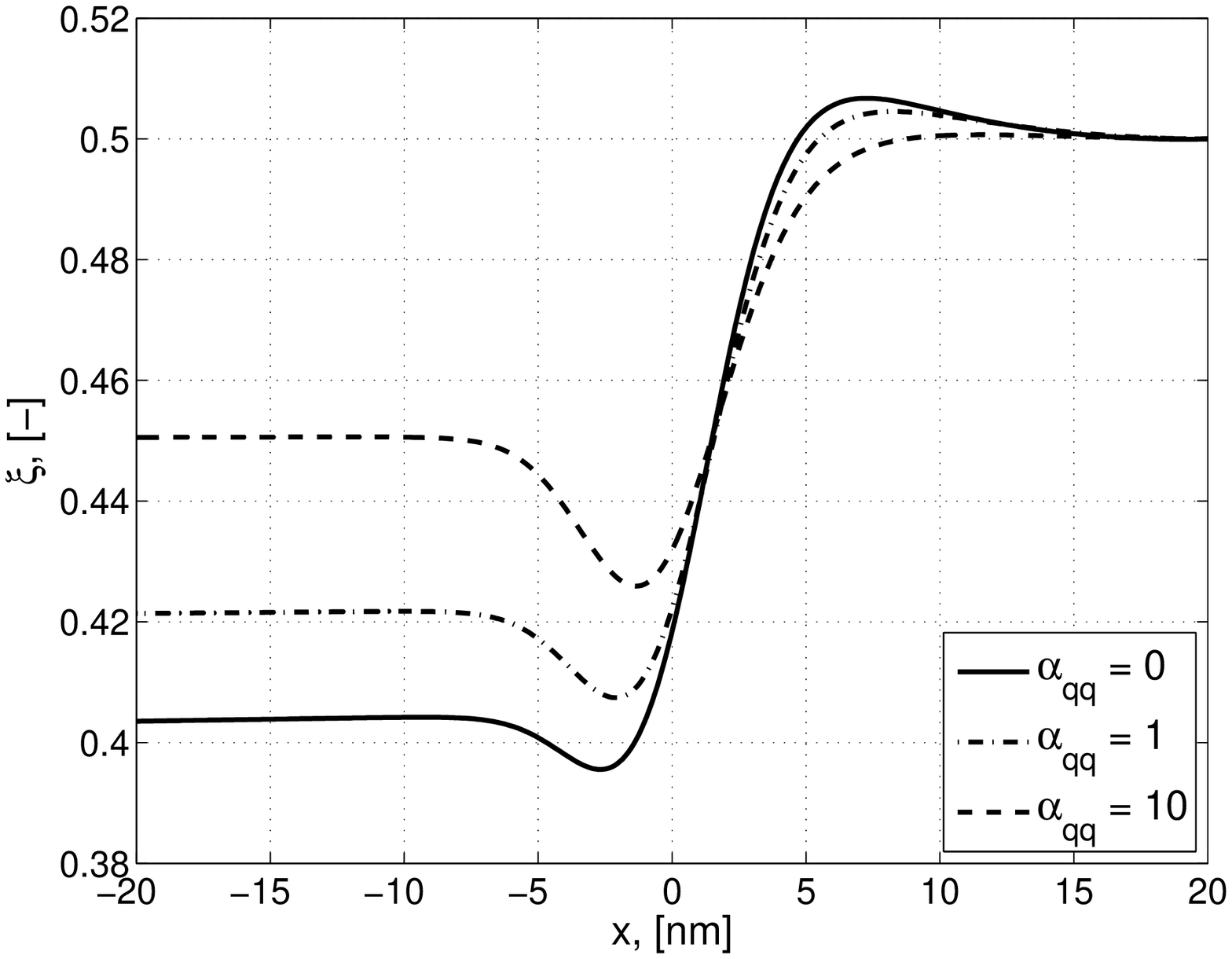}
\caption{Molar fraction profile for different $\alpha_{qq}$ at $p=p_{eq}$, $T_{g}=T_{eq}$ and $T_{\ell}=1.05 T_{eq}$}\label{Fig_xi-Aqq-p=pe_Tg=Te_Tl=1,05Te}
\includegraphics[scale=\scaleprofile]{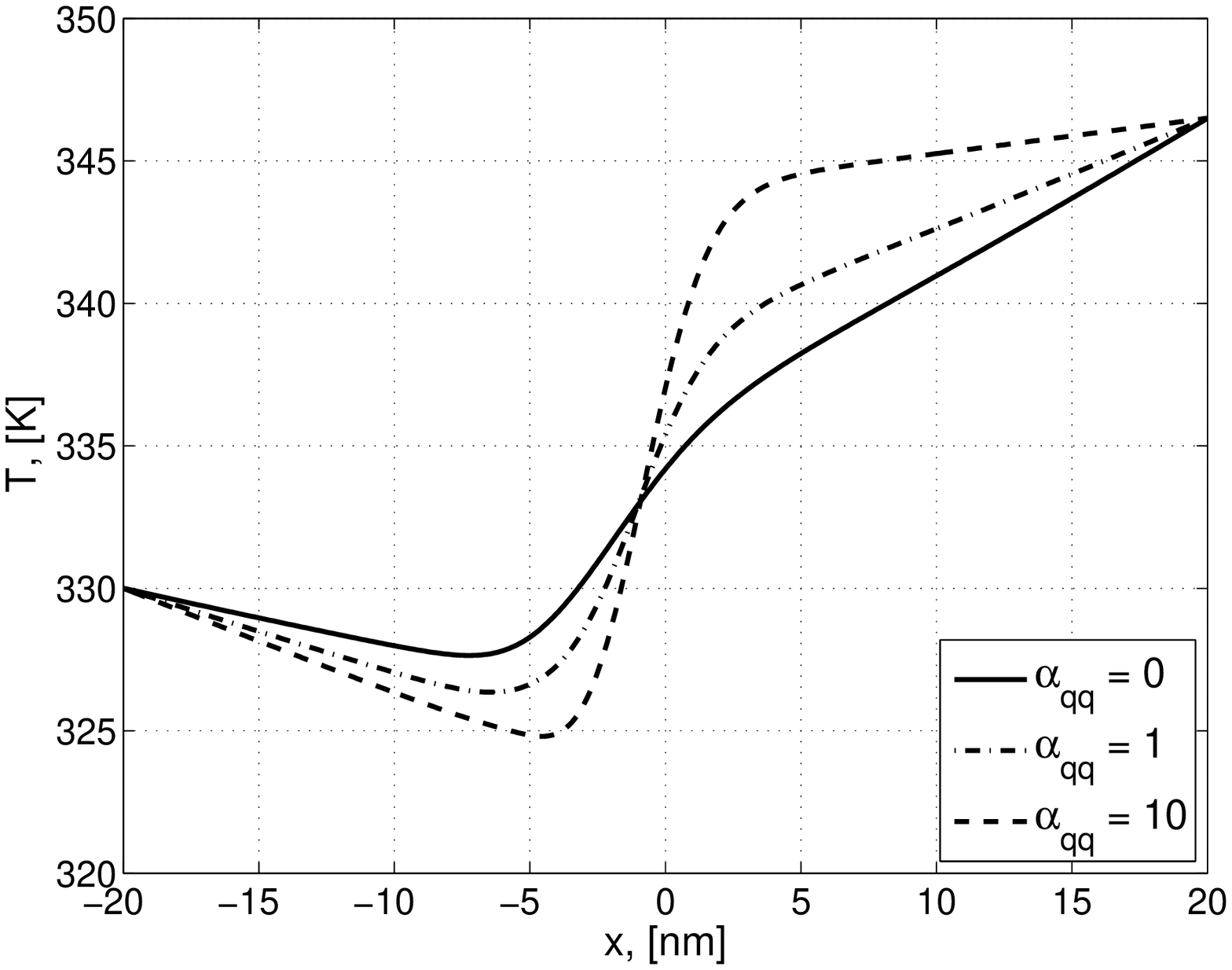}
\caption{Temperature profile for different $\alpha_{qq}$ at $p=p_{eq}$, $T_{g}=T_{eq}$ and $T_{\ell}=1.05 T_{eq}$}\label{Fig_T-Aqq-p=pe_Tg=Te_Tl=1,05Te}
\end{figure}

As fourth example we consider the $\alpha_{ii} \neq 0$ case when the vapor pressure and temperature are kept equal to the equilibrium values and the liquid temperature is 5 $\%$ higher then the
equilibrium value. The total concentration profile is still similar to \figr{Fig_c-Aqq-p=0,95pe_Tg=Tl=Te} and not given. \figsr{Fig_xi-Aii-p=pe_Tg=Te_Tl=1,05Te}{Fig_T-Aii-p=pe_Tg=Te_Tl=1,05Te}
give the mole fraction and temperature profiles. One can notice again the more dramatic behavior of the mole fraction and temperature profiles for big values of $\alpha_{ii}$.
%
%
\begin{figure}[hbt!]
\centering
\includegraphics[scale=\scaleprofile]{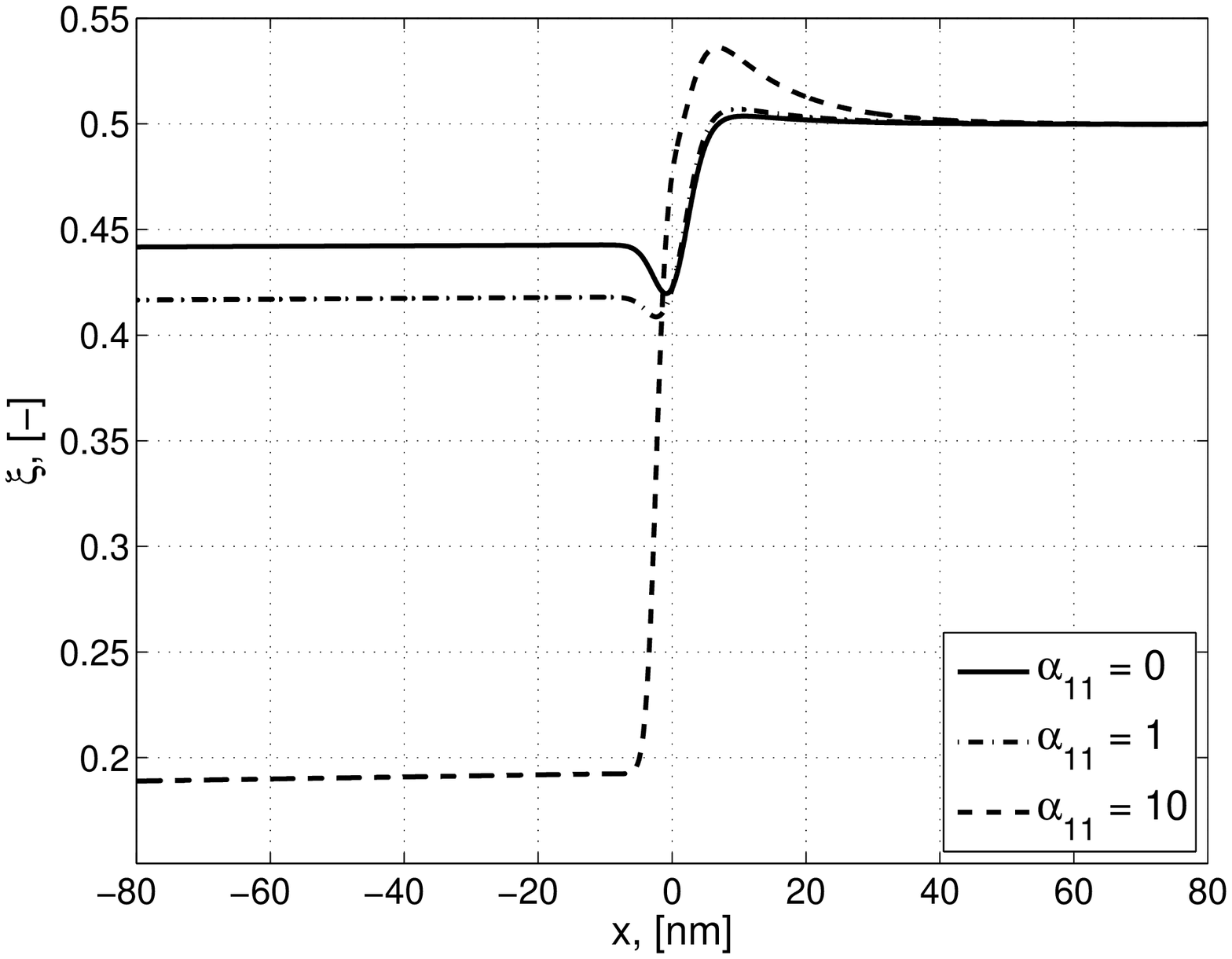}
\caption{Molar fraction profile for different $\alpha_{11}$ at $p=p_{eq}$, $T_{g}=T_{eq}$ and $T_{\ell}=1.05 T_{eq}$}\label{Fig_xi-Aii-p=pe_Tg=Te_Tl=1,05Te}
\includegraphics[scale=\scaleprofile]{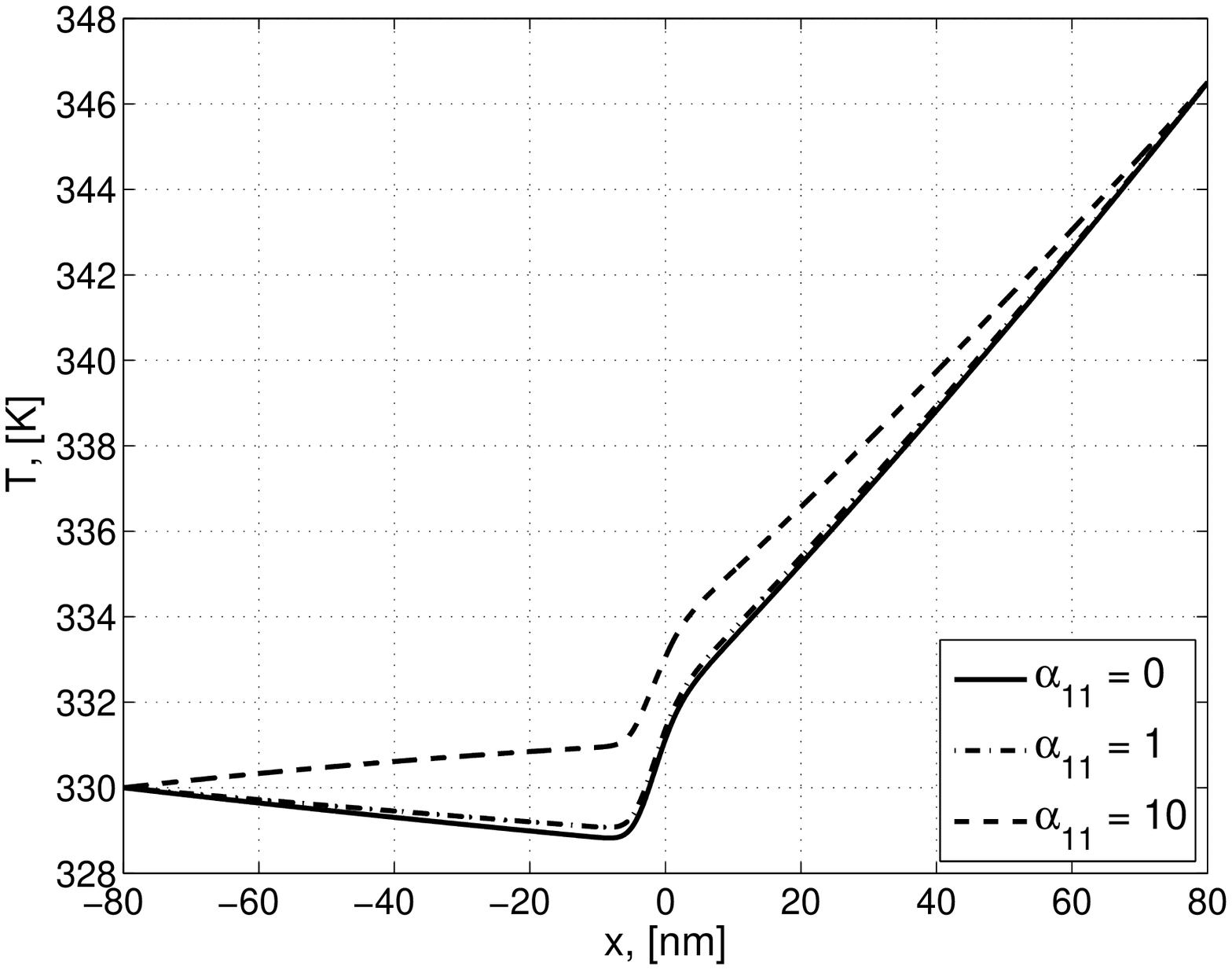}
\caption{Temperature profile for different $\alpha_{11}$ at $p=p_{eq}$, $T_{g}=T_{eq}$ and $T_{\ell}=1.05 T_{eq}$}\label{Fig_T-Aii-p=pe_Tg=Te_Tl=1,05Te}
\end{figure}

\begin{figure}[hbt!]
\centering
\includegraphics[scale=\scaleprofile]{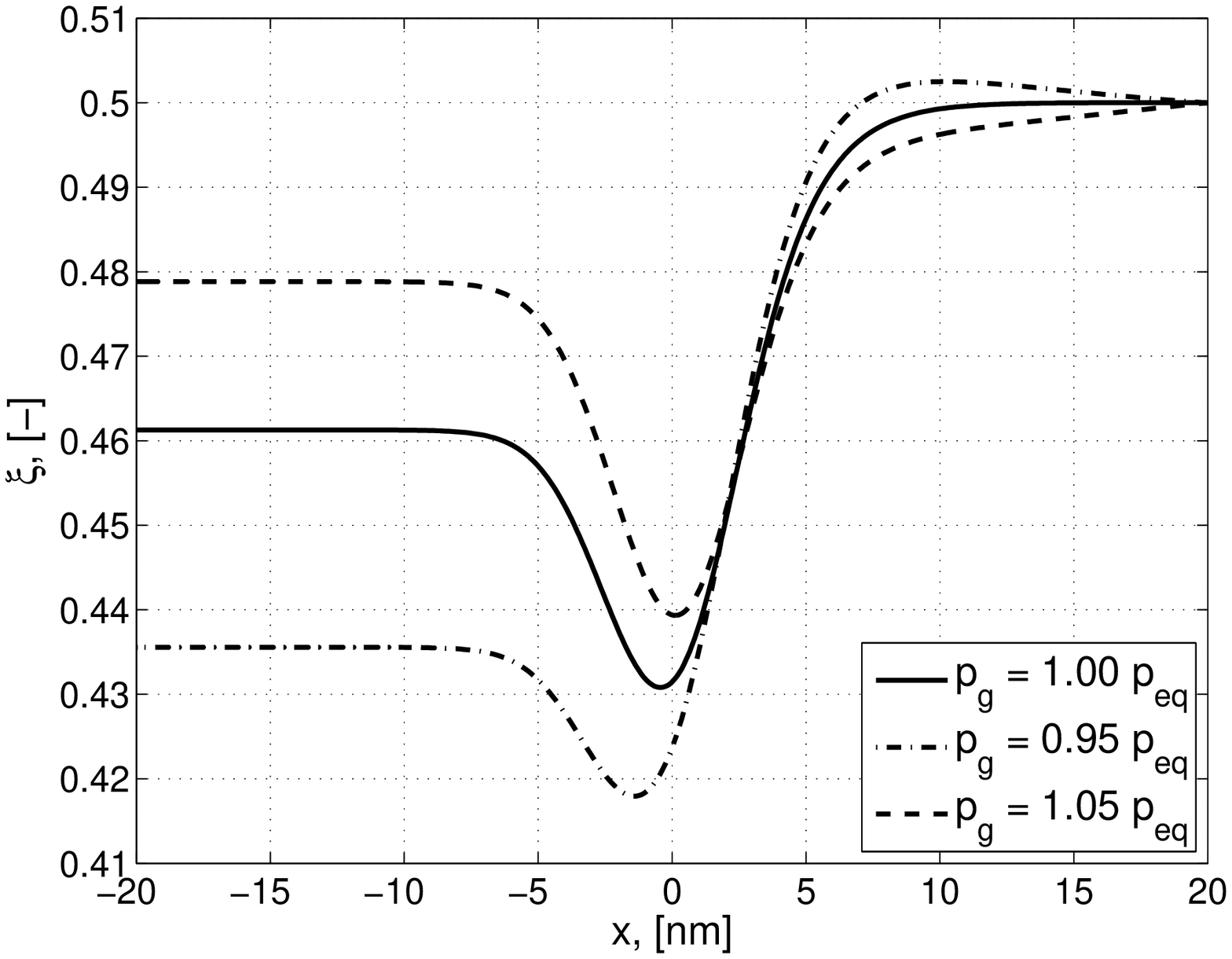}
\caption{Molar fraction profile for different $p$ at $T_{g}=T_{\ell}=T_{eq}$}\label{Fig_xi-p-Tg=Tl=Te_A=0}
\includegraphics[scale=\scaleprofile]{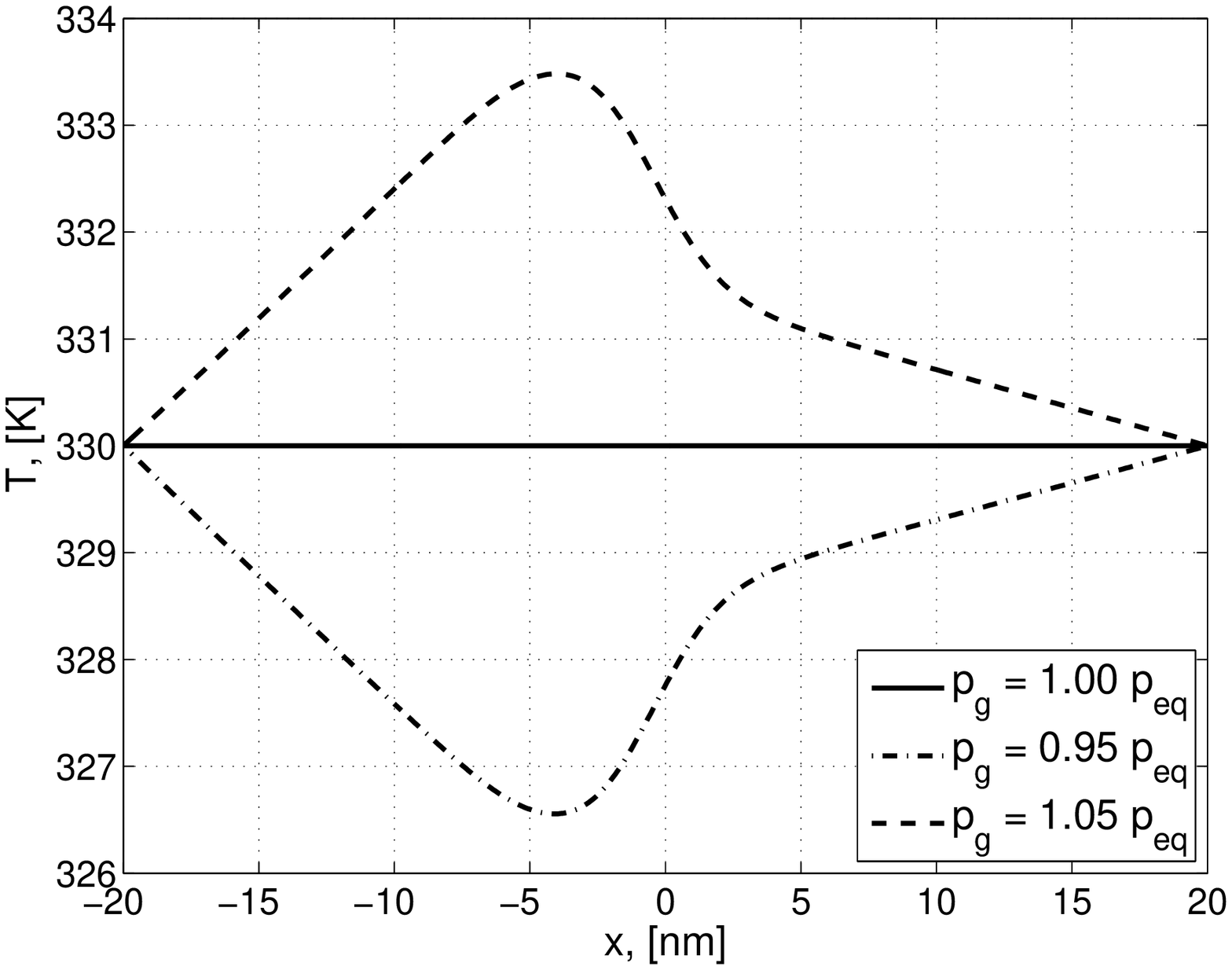}
\caption{Temperature profile for different $p$ at $T_{g}=T_{\ell}=T_{eq}$}\label{Fig_T-p-Tg=Tl=Te_A=0}
\end{figure}

Next we will discuss the influence of the pressure. We choose both $\alpha_{qq}$ and $\alpha_{11}$ equal to $1$ for this case. The total concentration profiles are again similar to
\figr{Fig_c-Aqq-p=0,95pe_Tg=Tl=Te} and not given. In \figsr{Fig_xi-p-Tg=Tl=Te_A=0}{Fig_T-p-Tg=Tl=Te_A=0} we give the mole fraction and the temperature profiles. The temperature profile goes down
for evaporation and up for condensation as expected. The mole fraction in the vapor rises a little bit for condensation and decreases for evaporation.
%
%

In the last case we consider the influence of the temperature on the liquid side. The total concentration profile is again similar to \figr{Fig_c-Aqq-p=0,95pe_Tg=Tl=Te} and not given. Lowering
(raising) $T_{\ell}$ gives evaporation (condensation). This lowers (rises) the temperature on the vapor side as expected. The mole fraction decreases (rises) about 5 $\%$ for condensation
(evaporation). This is the opposite of what happens in the previous case.
%
%
\begin{figure}[hbt!]
\centering
\includegraphics[scale=\scaleprofile]{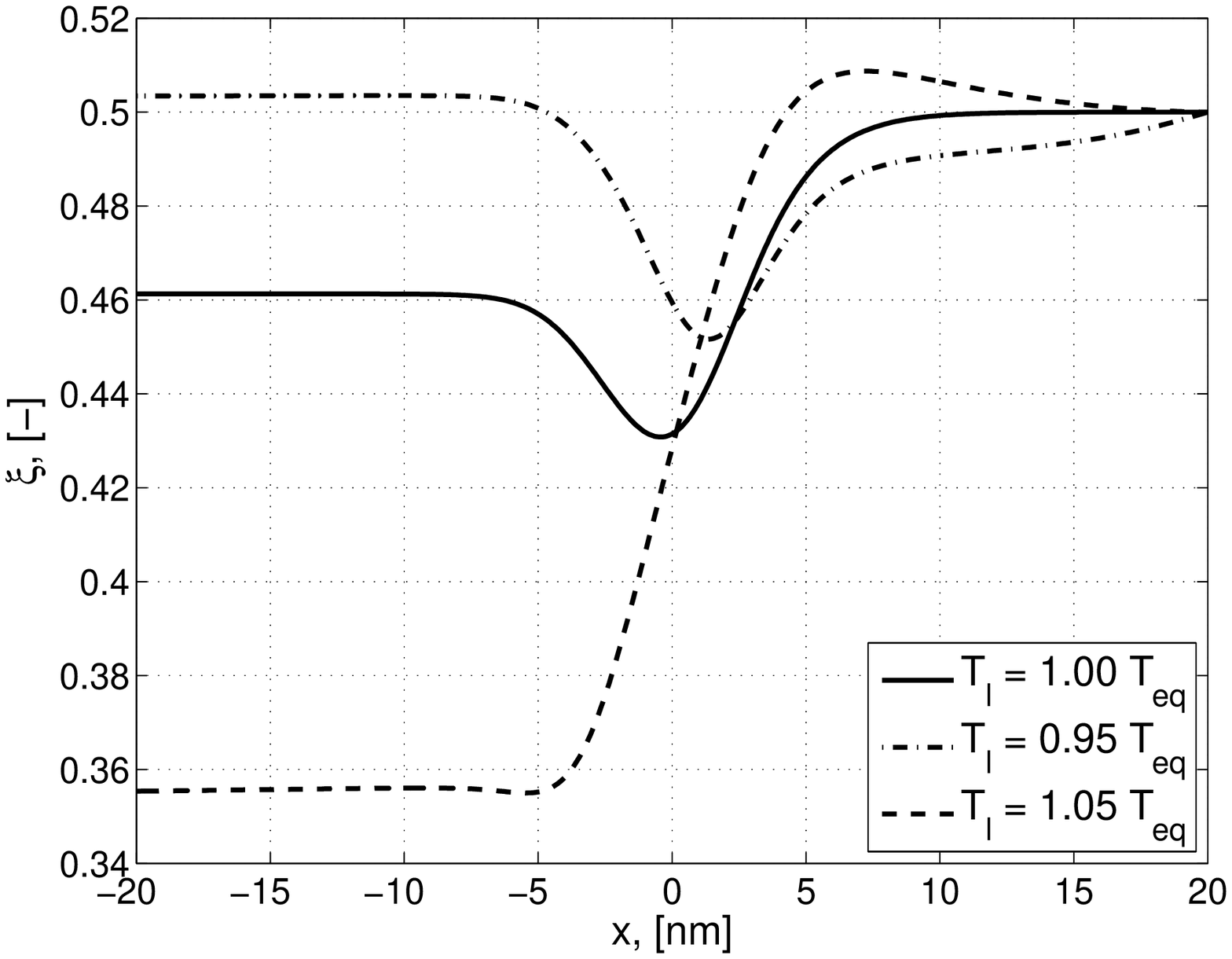}
\caption{Molar fraction profile for different $T_{\ell}$ at $T_{g}=T_{eq}$ and $p=p_{eq}$}\label{Fig_xi-Tl-p=pe_Tg=Te_A=0}
\includegraphics[scale=\scaleprofile]{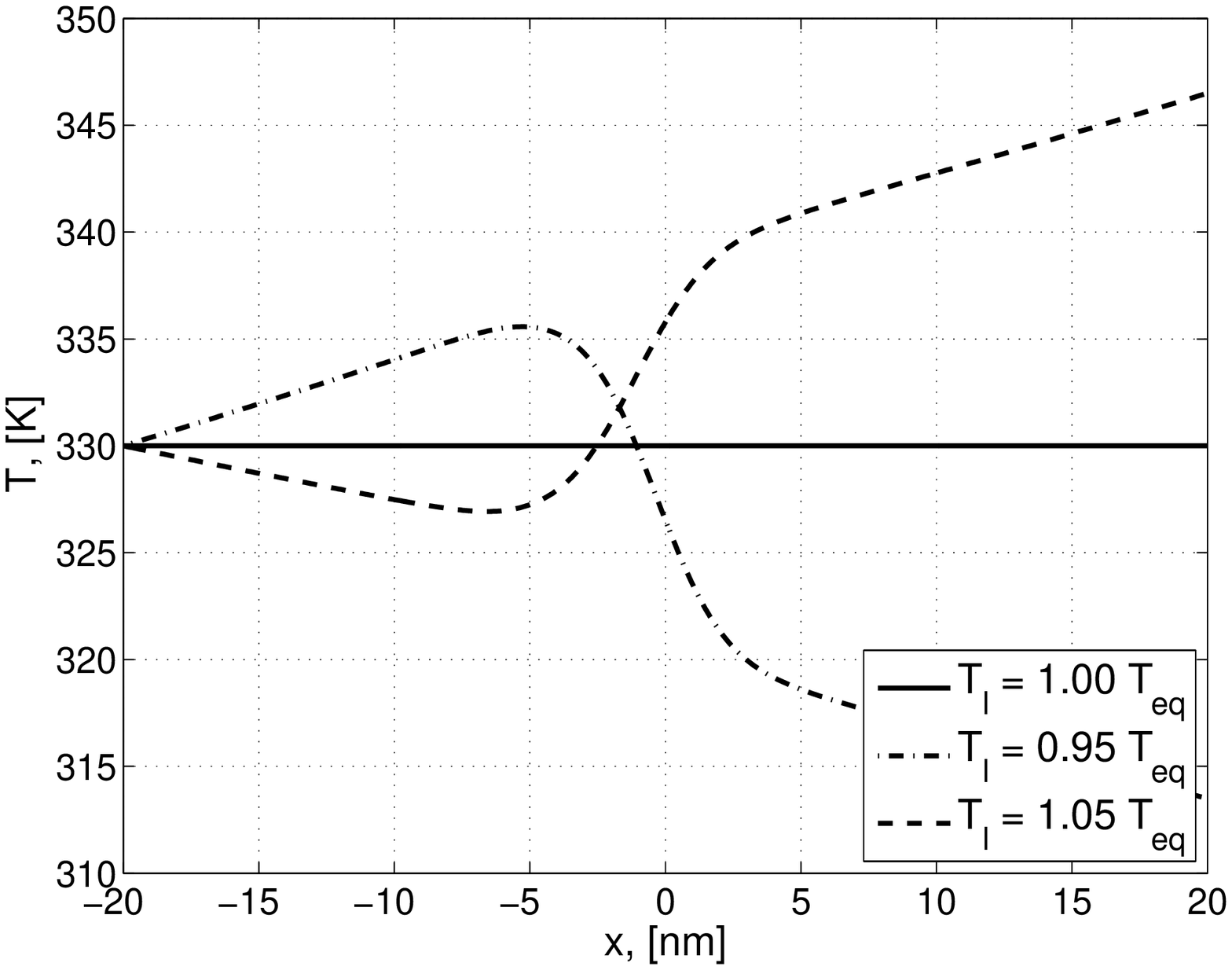}
\caption{Temperature profile for different $T_{\ell}$ at $T_{g}=T_{eq}$ and $p=p_{eq}$}\label{Fig_T-Tl-p=pe_Tg=Te_A=0}
\end{figure}

\section{Discussion and conclusions.}\label{sec/Discussion}

In this paper we have built the framework for the non-equilibrium gradient model for the surface. This require the following important steps. With the help of the equilibrium gradient model for
the mixtures, established in \cite{glav/gradient/eq/I} we were able to extend the thermodynamic description of the interfacial region to the non-equilibrium case. Explicit expressions for the
Gibbs relation and the pressure tensor were given. We found, that compared to homogeneous systems, the Gibbs relation for the interfacial region contains an additional term, proportional to the
divergence of the tension tensor. This tensor plays an important role in the surface and integration of it's perpendicular component gives the surface tension. Away from the surface this tensor
is equal to zero and we have the familiar Gibbs relation. For a single component system the Gibbs relation in the interfacial region reduces to the one given by \cite{bedeaux/vdW/I}.

For the non-equilibrium description we use the standard hydrodynamic equations for the so-called "one-fluid" model of the fluid. Together with the Gibbs relation and the balance equation for the
entropy density, we were then able to obtain explicit expressions for the entropy flux and the entropy production not only in the homogeneous phases but also in the interfacial region. This
identifies the conjugate thermodynamic forces and fluxes in the interfacial region. This made it possible to give the general force-flux relations in this region. The explicit form of these
equations depends on the symmetry of the system. We discuss why one can consider a fluid-fluid interface to be two-dimensional isotropic. Due to the Curie principle coupling occurs only between
2-dimensionally isotropic scalars, vectors and tensors of the same rank in such a system. The resulting force-flux relations in the interfacial region are accordingly simplified and given.
Interesting is that the components of the fluxes normal to the surface are scalar and couple therefore to reactions in the interfacial region. This is related to active transport, a phenomenon
of great importance. We do not explicitly consider reacting system in this paper, but may do this in a future publication. Having all these equations the non-equilibrium description of the
surface is complete.

We applied the description to the special case of stationary heat and mass transport through and into a surface. Transport along the surface is in many respects like a 2-dimensional analog of
flow in a homogeneous medium. The presence of the surface has more influence on transports in the normal direction. We find that there is for instance a strong effect of the surface on the
temperature and the concentration profiles. This effect increases when we increase the contribution proportional to the square gradient of the order parameter to the resistivity in the
interfacial region.

We conclude that the description we have given, using an extension of the square gradient model, will be a useful tool to study many details of the dynamics of evaporation and condensation in
multi-component systems. Non-equilibrium molecular dynamic simulations of evaporation and condensation could obtain density, mass fraction and temperature profiles. We expect the comparison of
these profiles with the present model to be very useful. In particular, this will give insight in the size and possible density, mass fraction and temperature dependence of the coefficient
$\alpha $. In this manner we expect the model to form a bridge between the microscopic description using non-equilibrium molecular dynamic simulations and the discrete macroscopic description
using the excess densities introduced by Gibbs \cite{gibbs/eqheter}. In a following paper we intend to investigate whether the discrete description satisfies the local equilibrium assumption for
an arbitrary choice of the dividing surface, a property which was verified for one-component systems \cite{bedeaux/vdW/II}. This would be a rather remarkable result, given the fact that the
continuous description does not obey this property. For a systematic development of the non-equilibrium thermodynamic description of surfaces this property is essential \cite{bedeaux/boundaryNE,
bedeaux/advchemphys}.

\appendix

\section{On the non-equilibrium Gibbs relation.}\label{sec/Gibbs}
For the specific Helmholtz energy the ordinary Gibbs relation in equilibrium was found to be
\begin{equation}\label{eq/Gibbs/01}
\delta f^{m}(s^{m}(\vR), v^{m}(\vR), \xi^{m}(\vR)) = -s^{m}(\vR)\,\delta T  + {\sum\limits_{i=1}^{n-1}{\psi^{m}_{i}\,\delta \xi^{m}_{i}(\vR)}} - p\,(\vR)\,\delta v^{m}(\vR)
\end{equation}%
Spatial Gibbs relations in equilibrium was found to be
\begin{equation}\label{eq/Gibbs/02}
\nabla f^{m}(\vR) = {\sum\limits_{i=1}^{n-1}{\psi_{i}^{m}\,\nabla \xi_{i}^{m}(\vR)}} - p\,(\vR)\,\nabla v^{m}(\vR) + v^{m}(\vR)\,{\PCDF{\gamma_{\alpha\beta}(\vR)}{\alpha}}
\end{equation}%

One may wonder why we use the Gibbs relations for the specific internal energy, not for the specific Helmholtz energy, to extend them to non-equilibrium analysis. Following the same procedure,
as in \ssecr{sec/Gradient/NonEquilibrium}, we can extend the Gibbs relations for the specific Helmholtz energy to non-equilibrium in the following way
\begin{equation}\label{eq/Gibbs/03}
{\frac{\partial f^{m}(\vRt)}{\partial t}} = - s^{m}(\vRt)\,{\frac{\partial T(\vRt)}{\partial t}} + {\sum\limits_{i=1}^{n-1}{\psi _{i}^{m}(\vRt)\,{\frac{\partial\xi_{i}^{m}(\vRt)}{\partial t}}}}
- p(\vRt)\,{\frac{\partial v^{m}(\vRt)}{\partial t}}
\end{equation}%
\begin{equation}\label{eq/Gibbs/04}
\nabla f^{m}(\vRt) = {\sum\limits_{i=1}^{n-1}{\psi _{i}^{m}(\vRt)\,\nabla \xi_{i}^{m}(\vRt)}} - p(\vRt)\,\nabla v^{m}(\vRt) + v^{m}(\vRt)\,{\PCDF{\gamma_{\alpha\beta}(\vRt)}{\alpha}}
\end{equation}%

For the ordinary Gibbs relations, like in homogeneous description, there is no preference in the thermodynamic potential. Provided \eqr{eq/Gradient/Eq/10}, equilibrium ordinary Gibbs relations
\eqr{eq/Gradient/Eq/11} and \eqr{eq/Gibbs/01} are equivalent. The non-equilibrium relation between these potentials
\begin{equation}\label{eq/Gibbs/05}
u^{m}(\vRt) = f^{m}(\vRt) + s^{m}(\vRt)\,T(\vRt) \\
\end{equation}
makes non-equilibrium ordinary Gibbs relations \eqr{eq/Gradient/NonEq/08a} and \eqr{eq/Gibbs/03} also to be equivalent.

The situation is different for the spatial Gibbs relations, however. Provided \eqr{eq/Gradient/Eq/10}, equilibrium spatial Gibbs relations \eqr{eq/Gradient/Eq/12} and \eqr{eq/Gibbs/02} are also
equivalent. The non-equilibrium relation \eqr{eq/Gibbs/05} between these potentials makes, however, non-equilibrium spatial Gibbs relations \eqr{eq/Gradient/NonEq/08b} and \eqr{eq/Gibbs/04} to
be \textit{not} equivalent. The reason for that is that the equilibrium spatial Gibbs relation for the specific Helmholtz energy does not contain the term, proportional to $\nabla T$, since
$\nabla T = 0$ in equilibrium. In non-equilibrium $\nabla T(\vRt) \neq 0$ and, as one can see from \eqr{eq/Gradient/NonEq/08b} and \eqr{eq/Gibbs/05}, $\nabla f^{m}(\vRt)$ contains such a term.

We see, that the Gibbs relations for the specific internal energy describe the system more adequately since they do not suffer from the unaccounted effect of possible temperature changes.
Because of this reason we should use the Gibbs relations for the specific internal energy, not the specific Helmholtz energy, to extend them to non-equilibrium.

\section{2D isotropic components in the 3D tensorial quantities}\label{sec/2Din3D}

As in \secr{sec/Curie2D} we shall use the special notation for the tensorial quantities of different order and different behavior in this section. Any tensorial quantity is denoted as
$Q^{(d\,\mathrm{r})}$. Here $d$ indicates the dimensionality of the space, in which the quantity is being considered, and can be either 3 or 2 here. $\mathrm{r}$ indicates the rank of the
tensorial quantity, and can be $\tS$ for scalar, $\tV$ for vectorial or $\tT$ for tensorial quantities. For example, $Q^{(2\tT)}$ indicates the 2-dimensional tensor, i.e. the quantity
$\big(\begin{smallmatrix}q_{11} & q_{12}\\ q_{21} & q_{22}\end{smallmatrix}\big)$, where $q_{ij}$ are numbers, and $Q^{(3\tV)}$ indicates the 3-dimensional vector, i.e. the quantity $\big(q_{1},
q_{2}, q_{3}\big)$, where $q_{i}$ are numbers. Scalars are the numbers irrespectively of the dimensionality of the space, so they will be denoted simply by $Q^{(\tS)}$.

Some quantities reveal the tensorial behavior of a some rank in $d$-dimensional space only under some specified transformations, while in general they don't. In this section we are interested
only in rotations around and reflections with respect to some constant vector $N^{(3\tV)}$ in 3-dimensional space. We will denote quantities which reveal the tensorial behavior of rank
$\mathrm{r}$ under these transformations by $Q^{(d\,\mathrm{r}\,_N)}$.

We show how in presence of the constant vector $N^{(3\tV)}$ one can split the tensorial quantity $Q^{(3\,\mathrm{r})}$ into a combination of the tensorial quantities $Q^{(2\,\mathrm{r}\,_N)}$.
Without loss of generality we will assume that $N^{\tV_3} = (1,\, 0,\, 0)$.

From 3D vector $V^{(3\tV)}$ one can construct the following quantities, which are linear in $V^{(3\tV)}$: one scalar quantity
$$V^{(\tS_N)} \equiv V^{(3\tV)} \spd N^{(3\tV)} = V^{(3\tV)}_1$$
and one vectorial quantity
$$V^{(3\tV_N)} \equiv V^{(3\tV)} - V^{(\tS_N)}\,N^{(3\tV)} = \big(0,\, V^{(3\tV)}_2,\, V^{(3\tV)}_3\big)$$
which is perpendicular to the $N^{(3\tV)}$. Denoting $V^{(2\tV_N)} \equiv \big(V^{(3\tV)}_2,\, V^{(3\tV)}_3\big)$ we can write that $V^{(3\tV_N)} = \big(0,\, V^{(2\tV_N)}\big)$. Thus,
\begin{equation}\label{eq/NonEquilibrium/Curie/02}
V^{(3\tV)} = V^{(\tS_N)}\,N^{(3\tV)} + V^{(3\tV_N)} = \big(V^{(\tS_N)},\,V^{(2\tV_N)}\big)
\end{equation}

From 3D tensor $T^{(3\tT)}$ one can construct the following quantities, which are linear in $T^{(3\tT)}$: 2 scalar quantity,
$$T^{(\tS)}_{0} \equiv \Tr\,T^{(3\tT)} = T^{(3\tT)}_{11}+T^{(3\tT)}_{22}+T^{(3\tT)}_{33}$$
and
$$T^{(\tS_N)}_{1} \equiv N^{(3\tV)} \spd T^{(3\tT)} \spd N^{(3\tV)} = T^{(3\tT)}_{11}$$
two vectorial quantities
$$T^{(3\tV_N)}_{l} \equiv N^{(3\tV)} \spd T^{(3\tT)} - T^{\tS_N}_{1}\,N^{(3\tV)} = \big(0,\, T^{(3\tT)}_{12},\, T^{(3\tT)}_{13}\big)$$
and
$$T^{(3\tV_N)}_{r} \equiv T^{(3\tT)} \spd N^{(3\tV)} - T^{\tS_N}_{1}\,N^{(3\tV)} = \big(0,\, T^{(3\tT)}_{21},\, T^{(3\tT)}_{31}\big)$$
(which are equal, if $T^{(3\tT)}$ is symmetric); and tensorial quantity
$$T^{(3\tT_N)} \equiv T^{(3\tT)} - T^{\tS_N}_{1}\,N^{(3\tV)}\,N^{(3\tV)} - T^{(3\tV_N)}_{l}\,N^{(3\tV)} - N^{(3\tV)}\,T^{(3\tV)}_{r} = \begin{pmatrix} \,0 & 0 & 0 \\ 0 & T^{(3\tT)}_{22} & T^{(3\tT)}_{23} \\ 0 & T^{(3\tT)}_{32} & T^{(3\tT)}_{33} \end{pmatrix}$$

Denoting
$$
\begin{array}{l}
T^{(2\tV_N)}_{l} \equiv \big(T^{(3\tT)}_{12},\, T^{(3\tT)}_{13}\big) \qquad  T^{(2\tT_N)} \equiv \begin{pmatrix} T^{(3\tT)}_{22} & T^{(3\tT)}_{23} \\ T^{(3\tT)}_{32} & T^{(3\tT)}_{33} \end{pmatrix}\\
T^{(2\tV_N)}_{r} \equiv \big(T^{(3\tT)}_{21},\, T^{(3\tT)}_{31}\big) \\
\end{array}
$$
we can write that
$$
\begin{array}{l}
T^{(3\tV_N)}_{l} = \big(0,\,T^{(2\tV_N)}_{l}\big)  \qquad  T^{(3\tT_N)} = \begin{pmatrix} 0 & 0 \\ 0 & T^{(2\tT_N)} \end{pmatrix}\\
T^{(3\tV_N)}_{r} = \big(0,\,T^{(2\tV_N)}_{r}\big) \\
\end{array}
$$
Thus\footnote{Note, that if the product of two tensorial quantities of rank $\mathrm{r} > 0$ is written without $\cdot\,$, it means that this is the product, not the internal product.}
\begin{equation}\label{eq/NonEquilibrium/Curie/05}
\begin{array}{ll}
T^{(3\tT)} &= T^{(\tS_N)}_{1}\,N^{(3\tV)}\,N^{(3\tV)} + T^{(3\tV_N)}_{l}\,N^{(3\tV)} + N^{(3\tV)}\,T^{(3\tV_N)}_{r} + T^{(3\tT_N)} \\
&= \begin{pmatrix} T^{(\tS_N)}_{1} & T^{(2\tV_N)}_{l} \\
T^{(2\tV_N)}_{r} & T^{(2\tT_N)} \end{pmatrix}
\end{array}
\end{equation}
Tensor $T^{(2\tT_N)}$ still contains the scalar part
$$T^{(\tS_N)}_{2} \equiv \Tr\,T^{(2\tT_N)} = T^{(3\tT)}_{32} + T^{(3\tT)}_{33}$$
which obeys the relation
\begin{equation}\label{eq/NonEquilibrium/Curie/06}
T^{(\tS)}_{0} = T^{(\tS_N)}_{1} + T^{(\tS_N)}_{2}
\end{equation}
Two of these three scalar quantities are linearly independent and one can use any pair. Since we want to reduce all the quantities to the form $Q^{(2\,\mathrm{r}\,_N)}$ we will use
$T^{(\tS_N)}_{1}$ and $T^{(\tS_N)}_{2}$ as independent pair. Introducing the traceless tensor
$$\traceless{T}^{(2\tT_N)} \equiv T^{(2\tT_N)} - \textstyle{1 \over 2}\,T^{(\tS_N)}_{2}\,U^{(2\tT)} = \begin{pmatrix} T^{(3\tT)}_{22}-\textstyle{1 \over 2}\,T^{(\tS_N)}_{2} & T^{(3\tT)}_{23} \\ T^{(3\tT)}_{32} & T^{(3\tT)}_{33}-\textstyle{1 \over 2}\,T^{(\tS_N)}_{2} \end{pmatrix} $$
and
$$\traceless{T}^{(3\tT_N)} \equiv T^{(3\tT_N)} - \textstyle{1 \over 2}\,T^{(\tS_N)}_{2}\,U^{(3\tT_N)} = \begin{pmatrix} \,0 & 0  \\ 0& \traceless{T}^{(2\tT_N)} \end{pmatrix} $$
where
$$U^{(2\tT)} \equiv \begin{pmatrix} 1 & 0 \\ 0 & 1 \end{pmatrix} \qquad U^{(3\tT_N)} \equiv \begin{pmatrix} \,0 & 0 \\ 0& U^{(2\tT)} \end{pmatrix}$$
we can write a 3D tensor as
\begin{equation}\label{eq/NonEquilibrium/Curie/07}
\begin{array}{ll}
T^{(3\tT)} &= T^{(\tS_N)}_{1}\,N^{(3\tV)}\,N^{(3\tV)} + T^{(3\tV_N)}_{l}\,N^{(3\tV)} + N^{(3\tV)}\,T^{(3\tV_N)}_{r} + {\textstyle{1 \over 2}}\,T^{(\tS_N)}_{2}\,U^{(3\tT_N)} + \traceless{T}^{(3\tT_N)} \\
&= \begin{pmatrix} T^{(\tS_N)}_{1} & T^{(2\tV_N)}_{l} \\ T^{(2\tV_N)}_{r} & {\textstyle{1 \over 2}}\,T^{(\tS_N)}_{2}\,U^{(2\tT)} + \traceless{T}^{(2\tT_N)}  \end{pmatrix}
\end{array}
\end{equation}
\section{Helmholtz energy of a mixture of ideal gases.}\label{sec/Helmholtz/Ideal}

According to \cite{ll5} the total Helmholtz energy of a homogeneous mixture of ideal gases is
\begin{equation}  \label{eq/IG/01}
F_{0,\,id}[\alTcxi] = - RT {\sum\limits_{k=1}^{n}{\nu_{k}\ln\Big({\frac{e\,\mathrm{w}_{k}(T)}{c_{k}\,N_{A}\,\Lambda_{k}^{3}(T)}}\Big)}}
\end{equation}
where $\nu_{k}$ is the number of moles and $c_{k}$ the molar density of component $k$. Furthermore $\Lambda_{k}$ is the thermal de Broglie wavelength and $\mathrm{w}_{k}$ is a characteristic sum
over the internal degrees of freedom of component $k$
\begin{equation}  \label{eq/IG/02}
\Lambda_{k}(T) \equiv \hbar\,N_{A}\sqrt{2\pi/M_{k}RT} \quad, \quad \mathrm{w}_{k}(T) \equiv {\sum\nolimits_{\ell}{\exp({-\varepsilon^{\,\ell}_{k}/k_{B}T})}}
\end{equation}
where $M_{k}$ is the molar mass of component $k$ and $\varepsilon^{\,\ell}_{k}$ are the energy levels of the internal degree of freedom of component $k$. If one describes the mixture using molar
specific variables the following equivalent expression is more useful
\begin{equation}\label{eq/IG/03}
F_{0,\,id}[\alTcxi] = -\nu RT\ln \Big({\frac{e\,\mathrm{w}(\alTxi)}{c\,N_{A}\Lambda^{3}(\alTxi)}}\Big) - RT{\sum\limits_{k=1}^{n}{\nu_{k}\ln
\Big({\frac{c}{c_{k}}}{\frac{\Lambda^{3}(T)}{\Lambda_{k}^{3}(\alTxi)}} {\frac{\mathrm{w}_{k}(T)}{\mathrm{w}(\alTxi)}}\Big)}}
\end{equation}
where $c$ is the total molar density of the mixture,
\begin{equation}\label{eq/IG/03}
\Lambda(\alTxi) \equiv \hbar N_{A}\sqrt{2\pi/M(\xi)RT}
\end{equation}
is the mixture's thermal de Broglie wavelength, $\mathrm{w}(\alTxi)$ a characteristic sum over all the internal degrees of freedom of the mixture and
\begin{equation}\label{eq/IG/04}
M(\xi) = {\sum\limits_{k=1}^{n}{\xi_{k}M_{k}}} = M_{n} + {\sum\limits_{k=1}^{n-1}{\xi_{k}(M_{k}-M_{n})}}%
\end{equation}
is the molar mass of the mixture. The exact expression for $\mathrm{w}(\alTxi)$, as well as expression for $\mathrm{w}_{k}(T)$, is determined by model approximation for the mixture.

The specific Helmholtz energy of a mixture of ideal gases then becomes
\begin{equation}\label{eq/IG/05}
f_{0,\,id}(\alTcxi) = -RT\ln\Big({\frac{e\,\mathrm{w}(\alTxi)}{c\,N_{A}\,\Lambda^{3}(\alTxi)}}\Big) -
RT{\sum\limits_{k=1}^{n}{\xi_{k}\ln\Big({\frac{1}{\xi_{k}}}\Big({\frac{M_{k}}{M(\xi)}}\Big)^{\!3/2}{\frac{\mathrm{w}_{k}(T)}{\mathrm{w}(\alTxi)}}\Big)}}
\end{equation}
Due to the spirit of a the one-fluid approach we have to equate the second term to $0$. Thus,
\begin{equation}\label{eq/IG/00}
f_{0,\,id}(\alTcxi) = -RT\,\ln\Big({\frac{e\,\mathrm{w}(\alTxi)}{c\,N_{A}\,\Lambda^{3}(\alTxi)}}\Big)
\end{equation}
where
\begin{equation}\label{eq/IG/06}
\mathrm{w}(\alTxi) = \exp\Big\{ {\sum\limits_{k=1}^{n}{\xi_{k}\ln\Big({\frac{1}{\xi_{k}}}\Big({\frac{M_{k}}{M(\xi)}}\Big)^{\!3/2}\mathrm{w}_{k}(T)\Big)}} \Big\}
\end{equation}
can be considered as a mixing rule for the $\mathrm{w}$. We note that \eqr{eq/IG/00} together with \eqr{eq/IG/06} does not impose any assumptions: it is nothing but \eqr{eq/IG/05} written with
the help of one-fluid terms.

\section{Symbols list}\label{sec/Symbols}

\begin{longtable}{c@{\quad--\quad}p{.8\linewidth}}
$\cdot$ & contraction sign\\
$:$ & double contraction sign\\
$\nrange$ & enumeration of all integers from $1$ to $n$\\
$\Tr$ & trace\\
$\traceless{T}$ & traceless part of a tensor\\
$\alpha, \beta$ & cartesian indices\\
$\perp$ & perpendicular direction to the surface\\
$\parallel$ & parallel direction to the surface\\
$\partial$ & partial differential\\
$d, \delta$ & differential\\
$\delta_{\alpha\beta}$ & Kroneker symbol\\
\end{longtable}

\begin{longtable}{l@{\quad}l@{\quad--\quad}p{.8\linewidth}}
$\nabla$ & [ 1/m ] & nabla operator\\
$\gamma_{\alpha\beta}$ & [ Pa ] & tension tensor\\
$\Kappa^{m}$ & [ J/kg ] & square gradient contribution\\
$\kappa^{m}$ & [ J m\tsup{5}/kg\tsup{2} ] & square gradient coefficients for molar units\\
$\kappa_{i}^{m}$ & [ J m/kg ] & -------- $>>$ -------- $>>$ --------\\
$\kappa_{ij}^{m}$ & [ J/m\tsup{3} ] & -------- $>>$ -------- $>>$ --------\\
$\kappa_{ij}^{v}$ & [ J m\tsup{5}/kg\tsup{2} ] & square gradient coefficients for volume units\\
$\Lambda$ & [ m ] & thermal de Broglie wavelength\\
$\mu_{n}^{m}$ & [ J/kg ] & mass chemical potential of component $n$\\
$\Pi$ & [ Pa ] & viscous pressure tensor\\
$\pi_{\alpha\beta}$ & [ Pa ] & -------- $>>$ -------- $>>$ --------\\
$\rho$ & [ kg/m\tsup{3} ] & mass density\\
$\rho_{i}$ & [ kg/m\tsup{3} ] & mass density of component $i$\\
$\sigma_{s}$ & [ J/(K m\tsup{3} s) ] & entropy production\\
$\sigma_{\alpha\beta}$ & [ Pa ] & total pressure tensor\\
$\tau^{m}$ & [ J/kg ] & kinetic energy density per unit of mass\\
$\phi^{m}$ & [ J/kg ] & potential energy density per unit of mass\\
$\xi_{i}$ & [ -- ] & molar fraction of component $i$\\
$\xi_{i}^{m}$ & [ -- ] & mass fraction of component $i$\\
$\psi_{i}^{m}$ & [ J/kg ] & reduced mass chemical potential  of component $i$\\
$A$ & [ kg m\tsup{5} / (mol s)\tsup{2} ] & van der Waals equation of state coefficient for a mixture\\
$a_{i}$ & [ kg m\tsup{5} / (mol s)\tsup{2} ] &  van der Waals equation of state coefficient for a pure component $i$\\
$B$ & [ m\tsup{3}/mol ] &  van der Waals equation of state coefficient for a mixture\\
$b_{i}$ & [ m\tsup{3}/mol ] &  van der Waals equation of state coefficient for a pure component $i$\\
$c$ & [ mol/m\tsup{3} ] & molar concentration\\
$e_{m}$ & [ J/kg ] & total energy density per unit of mass\\
$f^{m}$ & [ J/kg ] & Helmholtz energy density per unit of mass\\
$f_{0}^{m}$ & [ J/kg ] & homogeneous Helmholtz energy density per unit of mass\\
$\vg$ & [ m/s\tsup{2} ] & gravitational acceleration\\
$i, j, k$ & [ -- ] & component number\\
$\vJ_{e}$ & [ J/(m\tsup{2}s) ] & total energy flux\\
$\vJ_{k}^{m}$ & [ kg/(m\tsup{2}s) ] & total mass flux of component $k$\\
$\vJ_{q}$ & [ J/(m\tsup{2}s) ] & total heat flux\\
$L_{ab,xy}$ & [ $\sim$ ] & phenomenological conductivities\\
$N_{A}$ & [ 1/mol ] & Avogadro's number\\
$n$ & [ -- ] & number of components\\
$p$ & [ Pa ] & pressure\\
$Q^{(d\,r)}$ & [ $\sim$ ] & tensorial quantity of rank $r$ in $d$-dimensional space \\
$R$ & [ J/(K mol) ] & universal gas constant\\
$R_{ab,xy}$ & [ $\sim$ ] & phenomenological resistivities\\
$\vR$ & [ m ] & position\\
$s^{m}$ & [ J/(K kg) ] & entropy density per unit of mass\\
$T$ & [ K ] & temperature\\
$t$ & [ s ] & time\\
$u^{m}$ & [ J/kg ] & internal energy density per unit of mass\\
$v$ & [ m\tsup{3}/mol ] & molar volume\\
$v^{m}$ & [ m\tsup{3}/kg ] & volume per unit of mass\\
$\vvelocity$ & [ m/s ] & barycentric velocity\\
$\velocity_{i}$ & [ m/s ] & velocity of component $i$\\
$x$ & [ m ] & position\\
\end{longtable}

%

\bibliographystyle{unsrt}

\end{document}